\documentclass[trackchanges,twocolumn]{aastex701}

\usepackage{amssymb}
\usepackage{kotex}
\usepackage{graphicx}
\usepackage{dcolumn}
\usepackage{bm}
\usepackage{xcolor}
\usepackage{booktabs}
\usepackage{amsmath}

\usepackage{graphicx}
\usepackage{dcolumn}
\usepackage{bm}
\usepackage{aasmacros}
\usepackage{xcolor}
\usepackage{booktabs}

\usepackage{amssymb}
\usepackage{kotex}

\begin{document}

\title{The Sc, Ti, and V Abundance Discrepancy: Testing High-Mass IMF Variation and Massive-Star Rotation}

\author[0009-0009-8436-4314]{Soonchul Choi}
\email[show]{scchoi0211@ibs.re.kr}
\affiliation{%
Center for Exotic Nuclear Studies, Institute for Basic Science, Daejeon 34126, Korea
}%

\author[0000-0002-7440-1080]{Eda Gjergo}
\email[show]{eda.gjergo@gmail.com}
\affiliation{%
School of Astronomy and Space Science, Nanjing University, Nanjing 210093, China
}%
\affiliation{Key Laboratory of Modern Astronomy and Astrophysics (Nanjing University), \\Ministry of Education, Nanjing 210093, People's Republic of China}

\author[0000-0003-0787-1485]{Youngman Kim}
\email{ykim@ibs.re.kr}
\affiliation{%
Center for Exotic Nuclear Studies, Institute for Basic Science, Daejeon 34126, Korea
}%

\author[0000-0002-8619-359X]{Toshitaka Kajino}
\email{kajino@buaa.edu.cn}
\affiliation{School of Physics, and International Research Center for Big-Bang cosmology and Element Genesis, Beihang University, Beijing 100083, China
}
\affiliation{National Astronomical Observatory of Japan, 2-21-1 Osawa, Mitaka, Tokyo 181-8588, Japan
}%
\affiliation{Graduate School of Science, The University of Tokyo, Tokyo 113-0033, Japan}

\begin{abstract}

Scandium, titanium, and vanadium can be synthesized primarily in massive stars. Yet many of the current Galactic chemical evolution models under-produce these elements at early epochs.
Motivated by evidence that the initial mass function 
varied in the past on the Galactic disc, we examine how assumptions about massive-star rotation and the initial mass function affect the inferred evolution of Sc, Ti, and V. We compute a grid of one-zone Galactic chemical evolution models that varies the initial rotational velocity of massive stars and the high-mass slope of the initial mass function. We compare the resulting [X/Fe] vs [Fe/H] for X= Sc, Ti, and V tracks and cross-element correlations with Galactic abundance data.
We find that adopting rotating massive-star yields with an initial rotational velocity of  300 km/s brings the model trends closer to metal-poor observations, especially for halo stars ([Fe/H] $< -2$), and improves the joint behavior of Sc, Ti, and V. Variations of the high-mass slope of the initial mass function produce a secondary modulation. The remaining tensions, most apparent at solar to super-solar metallicities, motivate future work with a more complete treatment of the enrichment physics and model uncertainties.

\end{abstract}


\section{Introduction}

Scandium, titanium, and vanadium provide sensitive tests of massive-star nucleosynthesis in Galactic chemical evolution (GCE) models. Large spectroscopic surveys, including Gaia-ESO~\citep{gilmore2012gaia}, APOGEE~\citep{wilson2019apache}, and GALAH~\citep{buder2021galah+}, together with complementary programs~\citep{steinmetz2006radial, yanny2009segue, deng2012lamost}, have mapped the chemical structure of the Milky Way across its main stellar components. In parallel, high-resolution studies of nearby stars provide homogeneous abundances for iron-peak and $\alpha$-elements in the thin disc, thick disc, and halo~\citep{adibekyan12, battistini15, bensby14, bensby05, cayrel04, gratton03, ishigaki12, ishigaki13, jacobson15, lai08, reddy03, reggiani17, yong13}. These datasets include [X/Fe]–[Fe/H] relations for X = Sc, Ti, and V, allowing their enrichment histories to be traced from metal-poor halo stars to the Galactic disc, where
$
[\mathrm{X}/\mathrm{Y}] \equiv \log_{10}(\mathrm{X}/\mathrm{Y})-\log_{10}(\mathrm{X}/\mathrm{Y})_\odot .
$

Interpreting these abundance trends requires a GCE framework, because the measured ratios reflect the integrated effects of star formation, gas flows, stellar lifetimes, and nucleosynthetic yields. GCE predictions depend on several ingredients: the initial mass function~\citep[IMF][]{Kroupa+2026}, the star-formation rate~\citep{kennicutt1998star}, the rates and metallicities of gas infall and outflows~\citep[e.g.,][]{spitoni2021g,spitoni2011effects,pipino2004photochemical}, radial migration~\citep[e.g.,][]{Johnson+2021}, and stellar yields~\citep[e.g.,][and references therein]{Liang+23}. A consistent choice of these ingredients is therefore required when using GCE models to interpret the evolution of individual elements~\citep[see discussions in][]{Matteucci:2001}.

GCE models reproduce many broad features of Milky Way chemical evolution, including the main structure of the [$\alpha$/Fe]–[Fe/H] relation and the evolution of several light and iron-peak elements. Both one-zone~\citep[e.g.,][]{matteucci2021modelling} and multi-zone models~\citep[e.g.,][]{minchev2013chemodynamical} can match abundance trends in the solar neighborhood and other Galactic components. However, Sc, Ti, and V remain difficult to reproduce. Commonly used stellar-yield sets underproduce these elements relative to observations, both at low metallicity and near the solar value. For instance, \citet{Kobayashi:2020jes} show that standard core-collapse supernova yield grids do not produce sufficient amounts of Sc, Ti, and V. This discrepancy suggests that some of the relevant nucleosynthetic physics, or the population-level weighting of massive-star ejecta, is still incomplete in standard GCE models.

Several mechanisms may affect the production of these elements. Neutrino-induced nucleosynthesis in the innermost ejecta of core-collapse supernovae can enhance some odd-Z and iron-peak species through neutrino interactions~\citep{kobayashi2011evolution}. Aspherical or jet-like explosions can also modify Fe-peak yields through high-entropy, low-density ejecta~\citep{sneden2016iron}. These mechanisms may increase the predicted abundances of Sc and V, but their contributions and parameter dependences remain uncertain.

A closely related Milky Way GCE study is that of \citet{Prantzos2018chemical}, who explored the Limongi–Chieffi rotating massive-star yields in a one-infall model. Their Figure~13 shows abundance tracks for several elements, including Sc, Ti, and V, and provides an important reference point for our comparison. Building on this work, we examine how two population-level ingredients, the representative initial rotational velocity (IRV) of massive stars and the high-mass IMF slope, affect the Sc, Ti, and V trends and their cross-element correlations within the same one-zone framework.

The IMF sets the relative numbers of low- and high-mass stars and therefore controls the weighting of different enrichment channels. The classical IMF of \citet{salpeter1955luminosity} identifies a single power-law slope for massive stars, while \citet{Kroupa:2000iv} refined the low-mass end to match local star counts. Observational and theoretical work suggests that a top-heavy IMF, corresponding to a higher fraction of massive stars, may be relevant in environments with high star-formation rates or low metallicities~\citep[e.g.,][and references therein]{Kroupa+2026,gjergo2025mass}. Evidence from Milky Way stellar clusters also indicates that IMF parameters may vary with the physical conditions of the star-forming gas \citep[e.g.,][]{Li+2023}. Motivated by these results, we test whether variations in the high-mass IMF slope can alter the early production of Sc, Ti, and V, adopting the \citet{Kroupa:2000iv} IMF as the fiducial baseline.

Stellar rotation changes the internal structure and nucleosynthesis of massive stars. Rotation-induced mixing, driven by meridional circulation and shear instabilities, transports nuclear species between burning regions, modifies convective zones, and changes the temperatures and densities at which burning occurs. These effects can alter both the total yields and the relative production of individual isotopes. Rotation also affects the surface properties of massive stars, especially at subsolar metallicity, where enhanced mixing can enrich the surface and strengthen radiatively driven mass loss. Such effects have long been discussed in the context of primary $^{14}$N and $^{13}$C production at low metallicity~\citep{MeynetMaeder2002a,MeynetMaeder2002b,ChiappiniRomanoMatteucci2003,Chiappini2003,Chiappini2006}. For the present work, the rotating models of \citet{Limongi:2018qgr} are important because they predict enhanced yields of several odd-Z and iron-peak elements, including Sc, Ti, and V.

In this study, we use the one-zone GCE code GalCEM~\citep{Gjergo2023} with the rotating massive-star yields of \citet{Limongi:2018qgr} to evaluate how the high-mass IMF slope, $\alpha_3$, and the IRV of massive stars affect the chemical evolution of Sc, Ti, and V. We compare the resulting [X/Fe]–[Fe/H] tracks with observational data for thin-disc, thick-disc, and halo stars. We also examine the combined effects of IMF slope and stellar rotation, and test whether the same models can reproduce the observed cross-element correlations among Sc, Ti, and V. This allows us to assess which combinations of IMF and rotational properties provide the most plausible explanation for the observed behaviour of these elements in the Milky Way.

\section{GCE}\label{sec:method}
\subsection{Theoretical model}

GCE quantifies past enrichment by reconstructing the stellar birthrate history of a Galaxy and by applying nucleosynthetic yields to stars at the time of their death.
The stellar birthrate is a function of both star formation rate (SFR) and IMF \citep[see discussions in ][]{Gjergo2023, Matteucci2012}.
We examine how two key factors, the slope of the IMF at the massive stars and the IRV of massive stars, affect the production of Sc, Ti, and V in massive stars.

We adopt the canonical IMF \citep{Kroupa:2000iv}, representative of local Milky Way thin-disc stars and historically treated as universal for thin-disc main-sequence galaxies. It is defined as:
\begin{equation}
\phi(M_*) = 
\begin{cases} 
    2\phi_0 M_*^{-\alpha_1} & M_l \leq \frac{M_*}{M_\odot} \leq 0.5 \\
    \phi_0 M_*^{-\alpha_2} & 0.5 \leq \frac{M_*}{M_\odot} \leq 1.0 \\
    \phi_0 M_*^{-\alpha_3} & 1.0 \leq \frac{M_*}{M_\odot} \leq M_u
\end{cases}
\end{equation}
where $M_u = 120\;M_\odot$ and $M_l = 0.08\;M_\odot$ denote the upper and lower mass limits and $M_\odot$ is the solar mass, respectively.
The factor of 2 in the lowest-mass bin ($2\phi M_*^{-\alpha_1}$) enforces continuity at $0.5 \; M_{\odot}$.
The canonical slopes of the power~law are  $\alpha_1=1.3$, $\alpha_2=2.3$, and $\alpha_3=2.3$ \cite[see also][]{Kroupa+2026}. 
The normalization constant, $\phi_0$, is chosen such that \citep{MatteucciGreggio1986}
\begin{equation}\label{eq:IMFnorm}
    \int_{M_l}^{M_u} \; M_*\,\phi\left(M_*\right)\,\mathrm{d}M_* = 1 \, .
\end{equation} 
In practice, this normalization defines the IMF per unit stellar mass formed, so that the SFR can be distributed over the stellar-mass grid.

Massive stars, defined here as having masses above $M_*>10\;M_\odot$, are the primary contributors to the synthesis of iron-peak elements such as Sc, Ti, and V via core-collapse supernovae (CCSNe). A smaller contribution comes from asymptotic giant branch (AGB) stars. Their progenitors span masses between 0.8 and 8~$M_{\odot}$.
Both of these sources reside primarily in the $\alpha_3$ range of the IMF. Therefore, whether the IMF is canonical, top-heavy or top-light, the $\alpha_3$ slopes affect how many massive stars will be generated, and consequently how many of them will contribute to the enrichment of Sc, Ti, and V elements.
\cite{Kroupa:2000iv} suggests that $\alpha_3$ may vary within the range $\alpha_3=2.3\pm0.7$. To explore this effect, we consider three representative values: $\alpha_3=1.6$, $2.3$, and $3.0$.

\begin{figure}[!ht]
    \centering
    \includegraphics[width=\columnwidth]{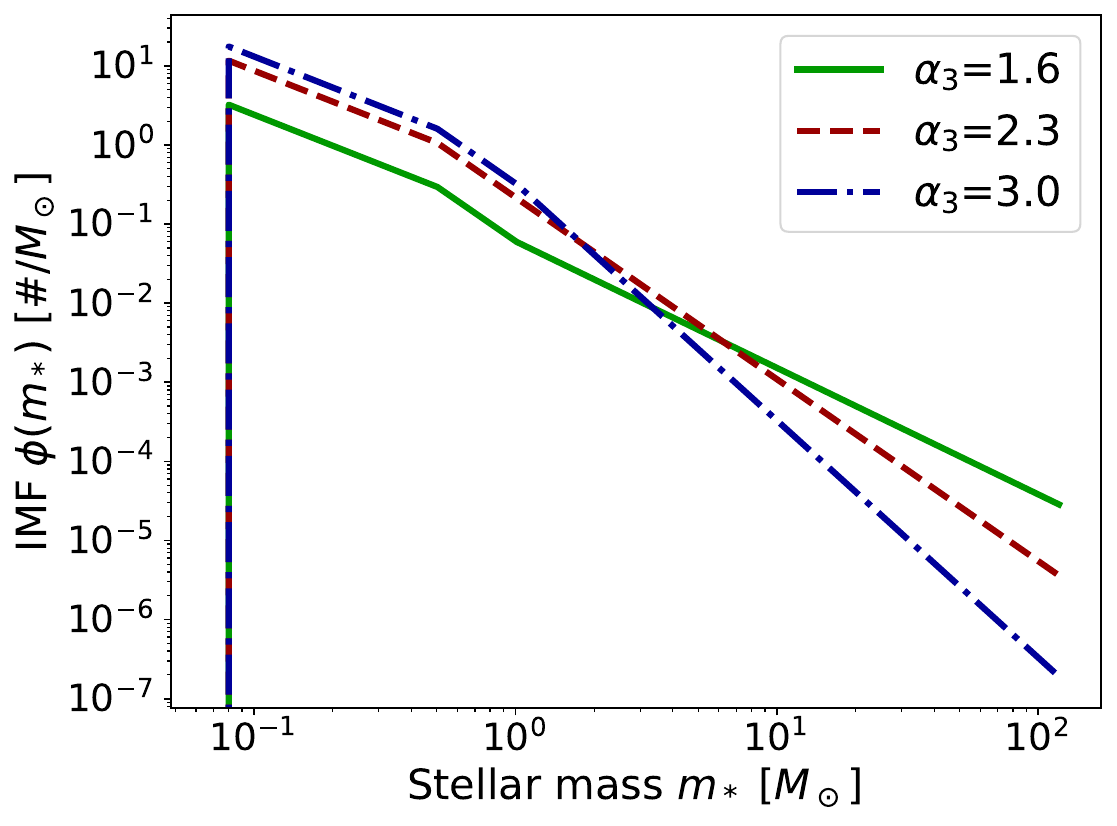}
    \caption{The initial mass functions (IMFs) for different high-mass slopes $\alpha_3 = 1.6$, 2.3, and 3.0. The IMF is normalized to unity (Eq.~\ref{eq:IMFnorm}). The canonical IMF slope for $m_* > 1 M_{\odot}$ \citep{Kroupa:2000iv} is $2.3$. The case with $\alpha_3=1.6$ corresponds to a top-heavy IMF and $\alpha_3=3.0$ to a top-light IMF.}
    \label{fig:imfs}
\end{figure}

Figure~\ref{fig:imfs} presents the IMF for different high-mass slopes, $\alpha_3 = 1.6$, $2.3$, and $3.0$, over the stellar mass range of $0.08$ to $120 M_\odot$. The normalization is chosen such that the integral of the mass-weighted IMF is unity (Eq.~\ref{eq:IMFnorm}).
As $\alpha_3$ increases, the IMF becomes steeper at the massive star, resulting in a lower fraction of massive stars.
Since these stars are the primary sites for the synthesis of elements such as Sc, Ti, and V, variations in $\alpha_3$ can influence their integrated yields in GCE models.

\begin{figure*}[ht!]
    \centering
    \includegraphics[width=0.49\textwidth]{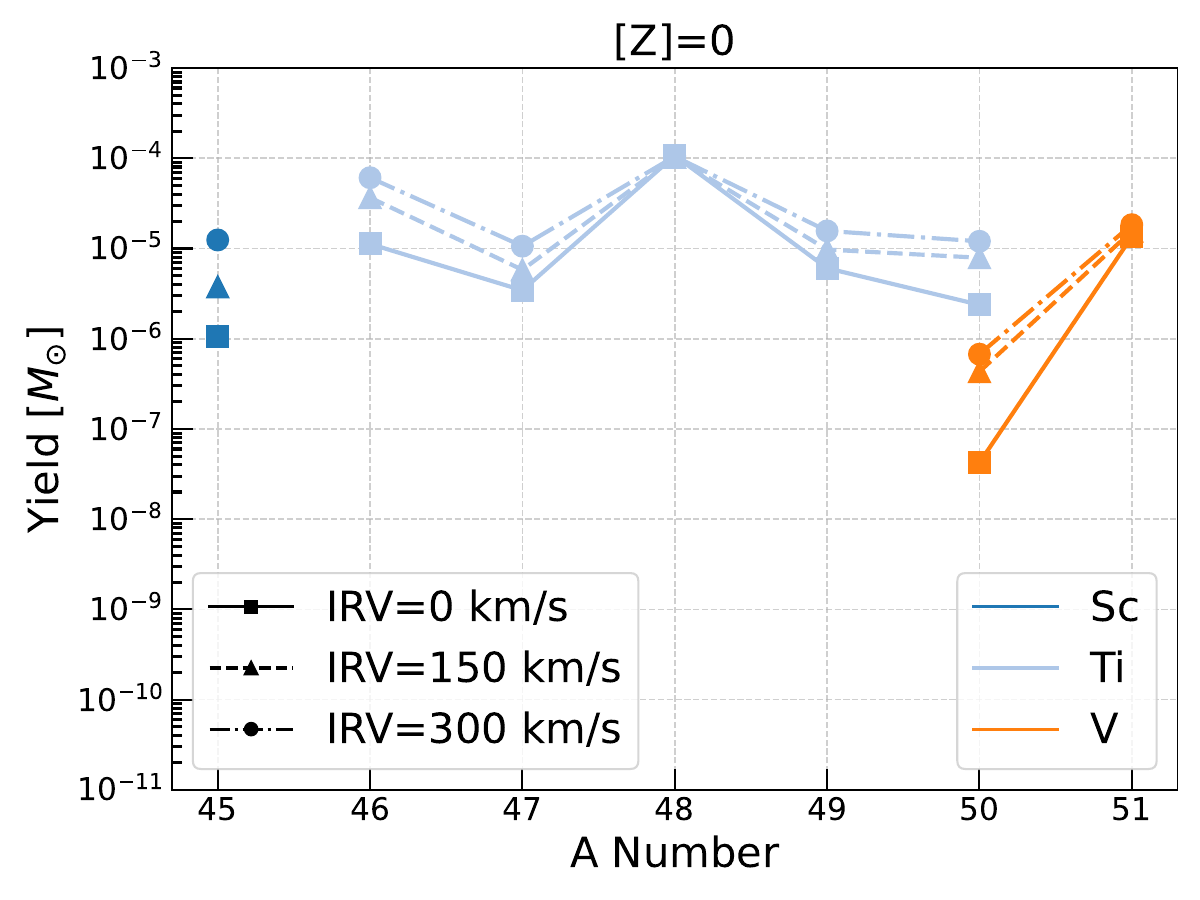}
    \includegraphics[width=0.49\textwidth]{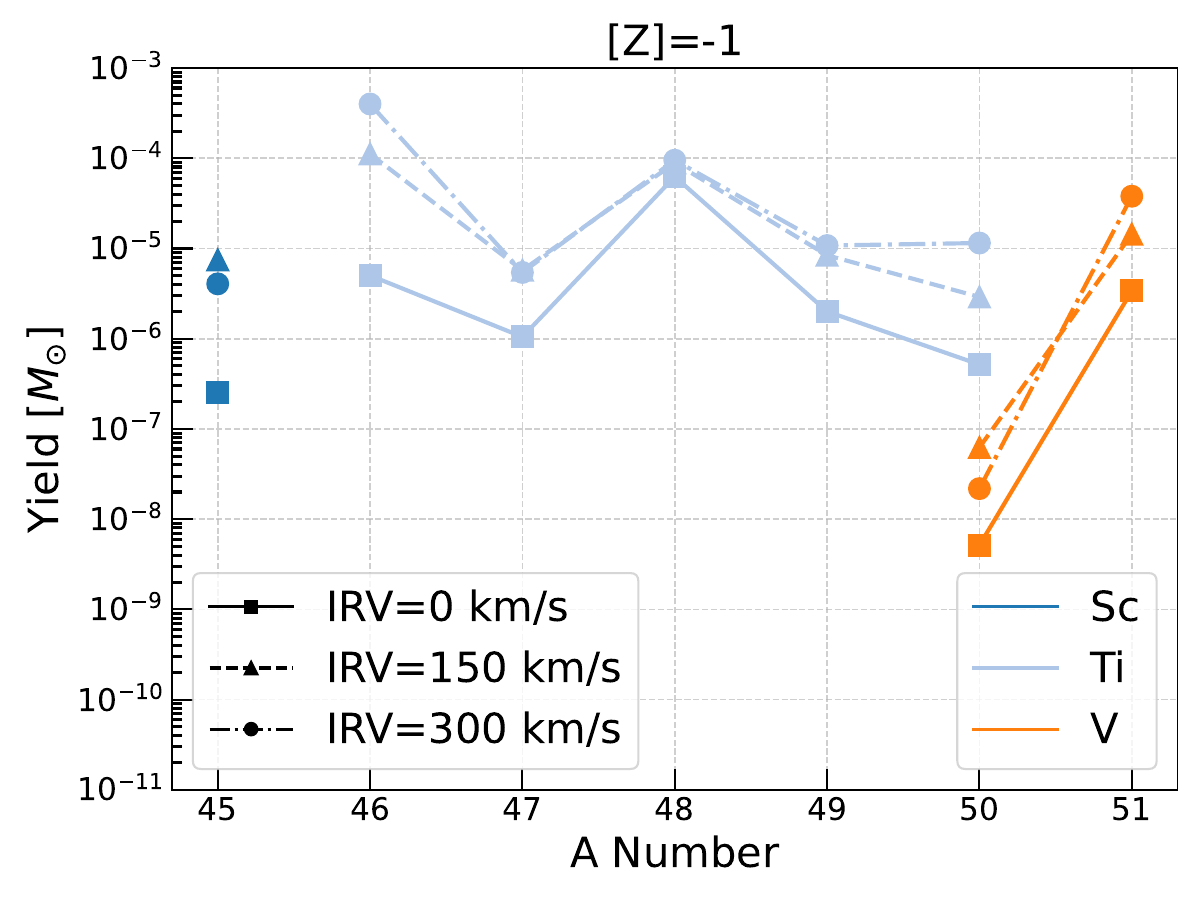}
    \includegraphics[width=0.49\textwidth]{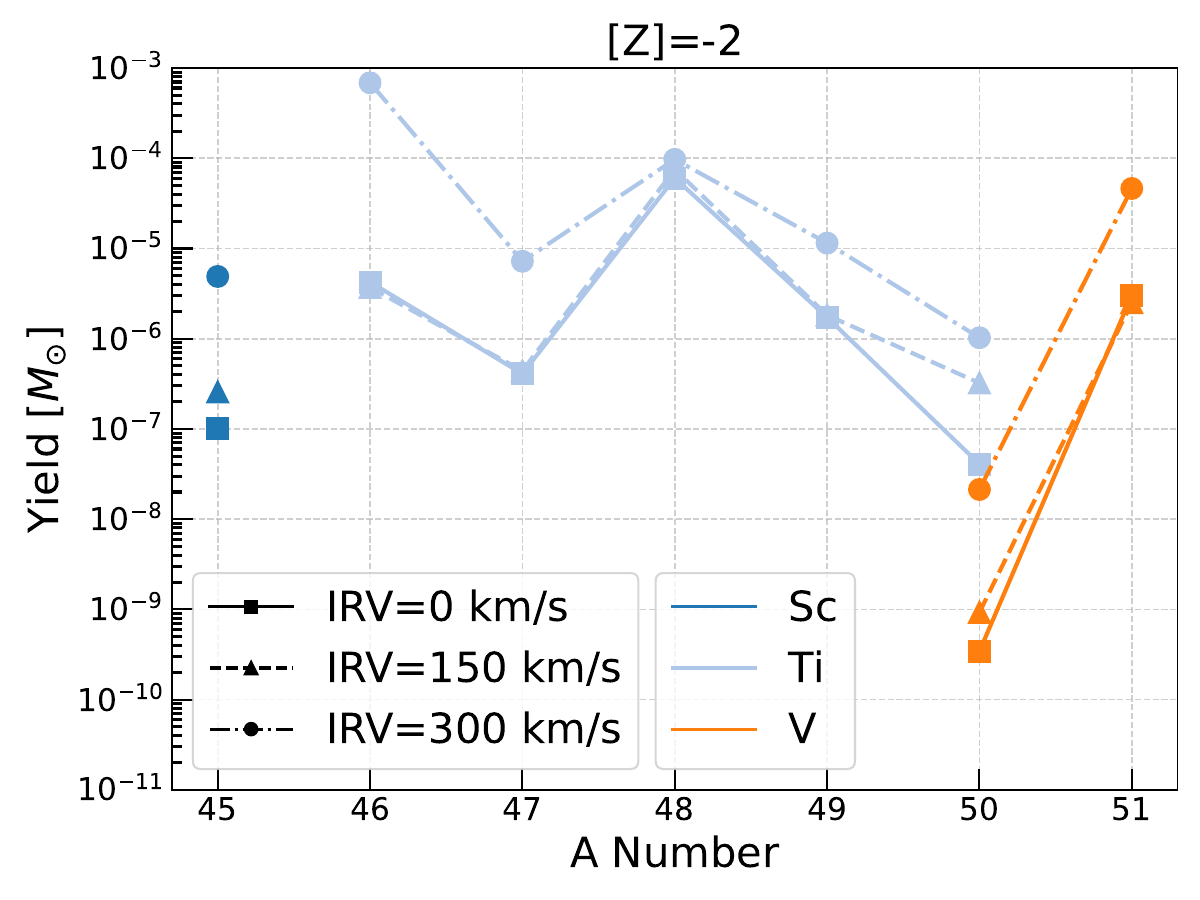}
    \includegraphics[width=0.49\textwidth]{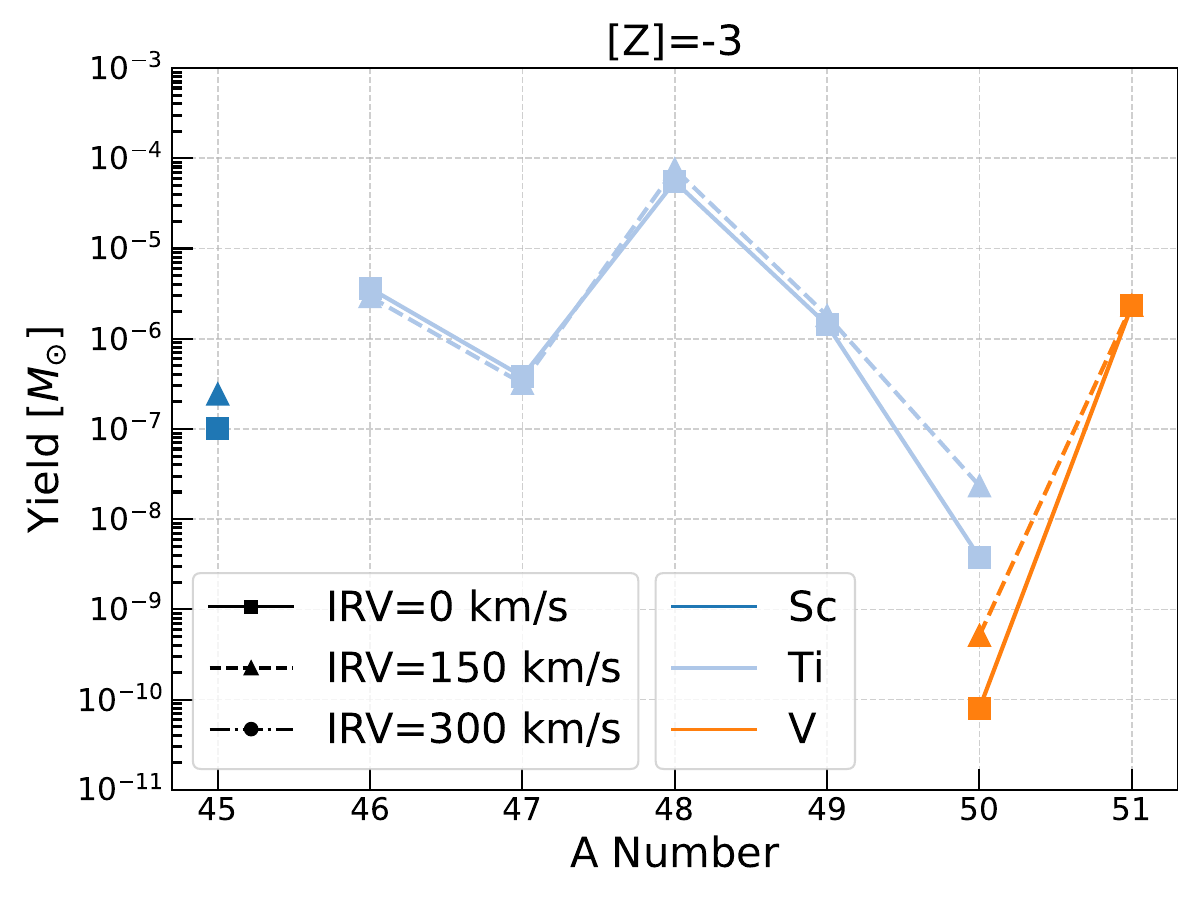}
    \caption{Isotope yields of Sc, Ti, and V from a $13 M_\odot$ star at various metallicities ($[Z]=\log_{10}(Z/Z_{\odot})$). The yields are taken from \cite{Limongi:2018qgr}. Different markers refer to different initial rotational velocities (IRV). The panels correspond to the following metallicities: $[Z]=0$ (top-left), $[Z]=-1$ (top-right), $[Z]=-2$ (bottom-left), and $[Z]=-3$ (bottom-right). For visual clarity, isotopes of the same element are connected by lines. Sc has a single stable isotope (dark blue). V has two (orange), and Ti has five (light blue). The IRV values considered are 0 km/s (solid line, square marker), 150 km/s (dashed line, triangle marker), and 300 km/s (dot-dashed line, circle marker). }
    \label{fig:yield_dep}
\end{figure*}

Figure~\ref{fig:yield_dep} reports the theoretical nucleosynthesis yields of Sc, Ti, and V for a $13$ $M_\odot$ massive star at various metallicities [$Z$], where $[Z]\equiv\log_{10}(Z/Z_{\odot})$, based on the models from \cite{Limongi:2018qgr}. The four panels correspond to $[Z]=0$, $[Z]=-1$, $[Z]=-2$, and $[Z]=-3$, and show how element yields vary with the initial rotational velocities of $0\;\mathrm{km/s}$, $150\;\mathrm{km/s}$, and $300\;\mathrm{km/s}$.

In general, the yields of Ti increase with IRV across all metallicities, indicating that stellar rotation enhances the production of these elements in massive stars. 
Sc and V, however, show a more complex IRV dependence.
Although they generally follow a similar increasing trend with IRV, an exception is observed at $[Z]=-1$, where the yield does not increase monotonically.
This suggests that the rotational dependence of Sc and V synthesis may involve additional factors, such as changes in burning conditions or convective mixing.

Additionally, metallicity also plays a significant role in determining absolute yields.
At higher $Z$, the yields of all three elements tend to be enhanced, reflecting the greater availability of seed nuclei for nucleosynthesis and the sensitivity of stellar structure to initial composition.

The chemical evolution calculations were performed using the publicly available GalCEM code \citep{Gjergo2023}, which tracks the evolution of elemental abundances over cosmic time by incorporating various stellar yields for CCSNe, AGB stars, and SNe Ia \citep{Cristallo:2015ica,Iwamoto:2000as,Limongi:2018qgr}, IMF assumptions \citep{Kroupa:2000iv}, and star formation prescriptions \citep{Portinari:1997bh}. In the present paper, this one-zone setup provides a baseline framework for isolating how the high-mass IMF slope and the representative massive-star IRV affect Sc, Ti, and V. The gas reservoir is assumed to be well mixed, with no outflow, and an exponentially decaying single infall, with timescale of 7~Gyr. Only $\alpha_3$ and the CCSN-yield choice tied to the adopted IRV are changed.

The massive-star yields are taken directly from the published tables of \cite{Limongi:2018qgr}. Their stellar models include rotation-induced transport by meridional circulation and shear instabilities, together with their adopted prescriptions for mass loss and explosion/remnant outcomes. In these models, rotational mixing can transport C and O produced in He-burning regions into H-burning layers, where primary $^{14}$N is produced and can later contribute to $^{22}$Ne production during He burning. This sequence changes the neutron excess and the pre-supernova structure, both of which are relevant for the synthesis of Sc, Ti, and V. The mass cut and black-hole formation criteria therefore follow those adopted in the published yield tables, and this inherited yield physics should be kept in mind when interpreting the resulting abundance trends. Likewise, each IRV case in our GCE runs is best regarded as an idealized case that isolates one representative massive-star rotation speed. A full birth-velocity distribution would be more realistic, but its dependence on stellar mass and metallicity is still uncertain. Altering the IRV across runs therefore provides a transparent way to quantify the sensitivity of the abundance trends to the adopted rotating-yield set.

\subsection{Observational data}
To compare our predictions with observed elemental abundances, we compiled a diverse set of observational datasets spanning stars in the thin disc, thick disc, and halo, as summarized in Table~\ref{tab:elemobs}. These datasets cover a wide range of metallicities and stellar populations, allowing us to compare our results with the observed abundances from various stars in the Milky Way.

\cite{adibekyan12} focused on [X/H] to characterize planet-host and non-host stars from the thin and thick discs.
\cite{battistini15} measured Sc, V, Mn, and Co abundances for a large sample of F- and G-type dwarfs in both the thin and thick discs.
\cite{bensby05,bensby14} provided chemical abundances for F- and G-type dwarfs and subgiants across thin and thick disc populations.
\cite{cayrel04} investigated very metal-poor halo giants selected from the ESO "First Stars" program.
\cite{cohen13} presented abundance patterns of extremely metal-poor (EMP) stars in the Galactic halo, including both carbon-enhanced metal-poor (CEMP) and non-CEMP populations.

\cite{gratton03} studied 150 subdwarfs and early subgiants in the halo and thin disc with accurate parallaxes and kinematic classifications.
\cite{jacobson15} analyzed 122 metal-poor stars in the halo, including a subset with [Fe/H]$\le-3.5$.
\cite{hinkel14} focused on stars in the thin disc, while \citep{ishigaki12,ishigaki13} provided elemental abundances for a kinematically selected sample from the thick disc and halo.
\cite{lai08} compiled detailed elemental abundances across evolutionary stages for halo stars with $-4\le$[Fe/H]$<-2$.
\cite{reddy03} combined samples from the thin discs to study elemental trends across Galactic components.
\cite{reggiani17} targeted halo stars and combined their results with literature data to study GCE across $-3.7<$[Fe/H]$<-0.4$.
\citep{yong13} focused on the metallicity distribution of CEMP stars, particularly those in the halo.
This compilation ensures broad coverage across metallicity and Galactic environments, enabling robust comparisons with model predictions. In the abundance figures below, point colours identify the literature subsamples used in the compilation rather than a one-to-one mapping onto Galactic component alone. Blue denotes studies containing both thin- and thick-disc stars, black denotes thin-disc-only samples, red denotes halo-only samples, and green denotes mixed thick-disc/halo samples.

\begin{table*}[ht!]
\footnotesize  
\centering
\caption{List of observational datasets included in Fig.~\ref{fig:alpha3_obs}.}
\begin{tabular}{l*{11}{c}}
\toprule
&Sc&Ti&V&  Where the stars are? \\ \midrule
\cite{adibekyan12}  & $\times$ & $\times$ & $\times$ & thick and thin disc\\
\cite{battistini15} & $\times$ & $\bigcirc$ & $\times$ & thick and thin disc\\
\cite{bensby05}     & $\bigcirc$ & $\times$ & $\bigcirc$ & thick and thin disc \\
\cite{bensby14}     & $\bigcirc$ & $\times$ & $\bigcirc$ & thick and thin disc \\
\cite{cayrel04}     & $\times$ & $\times$ & $\bigcirc$ & halo \\
\cite[no CEMP,][]{cohen13} & $\times$ & $\times$ & $\times$ & halo\\
\cite[CEMP,][]{cohen13}    & $\times$ & $\times$ & $\times$ & halo\\
\cite{gratton03} & $\times$ & $\times$ & $\times$ & halo\\
\cite{hinkel14}  & $\times$ & $\times$ & $\times$ & thin disc \\
\cite{ishigaki12,ishigaki13} & $\times$ & $\times$ & $\times$ & thick, halo\\
\cite{jacobson15}& $\times$ & $\times$ & $\bigcirc$ & halo\\
\cite{lai08}     & $\times$ & $\times$ & $\times$ & halo \\
\cite{reddy03}   & $\times$ & $\times$ & $\times$ & thin disc \\
\cite{reggiani17}& $\times$ & $\times$ & $\times$ & halo \\
\cite{yong13}    & $\times$ & $\times$ & $\bigcirc$ & halo \\
\bottomrule
\end{tabular}
   \label{tab:elemobs}
   \tablecomments{Each entry is marked with an $\times$ if the corresponding element (Sc, Ti, or V) was analyzed in the respective study, and with a $\bigcirc$ if it was not.
The processed public data is available for download at \url{https://github.com/egjergo/GalCEM/tree/main/galcem/input/observations/abund}.}
\end{table*}

\begin{figure*}[ht!]
    \centering
    \includegraphics[width=0.95\textwidth]{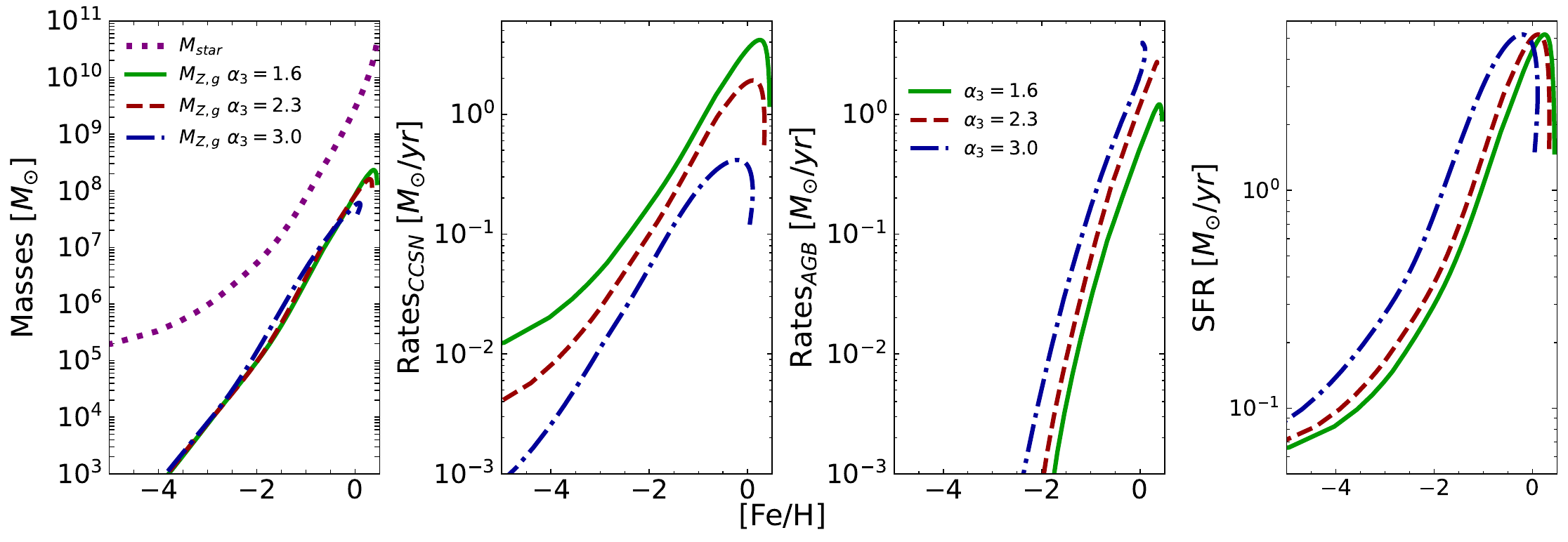}
    \caption{Four-panel comparison of the one-zone GCE histories for three high-mass IMF slopes, $\alpha_3=1.6$, 2.3, and 3.0, using the non-rotating massive-star yield set ($\mathrm{IRV}=0$ km/s). From left to right, the panels show the total stellar mass together with the metal-enriched gas mass, the CCSN mass-return rate, the AGB mass-return rate, and the star-formation rate. In the first panel, the dotted purple line denotes the total stellar mass, $M_\mathrm{star}$. The green solid, red dashed, and blue dot-dashed lines correspond to $\alpha_3=1.6$, 2.3, and 3.0, respectively. The mass-return rates and the SFR are expressed in $M_\odot$/yr.}
    \label{fig:alpha3_phys_com}
\end{figure*}
\begin{figure*}[ht!]
    \centering
    \includegraphics[width=0.95\textwidth]{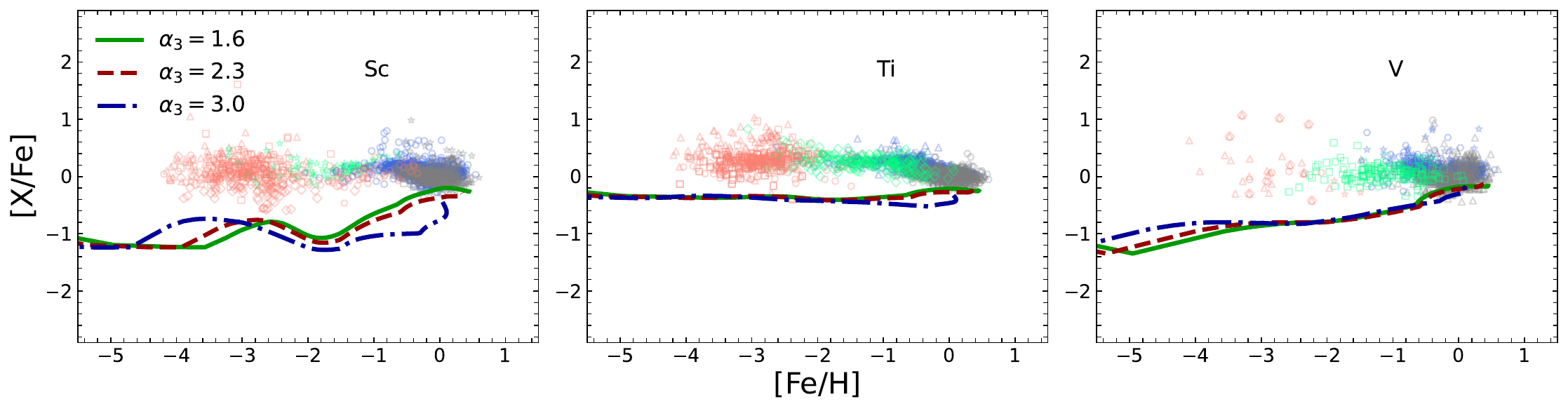}
    \caption{The abundance ratios [X/Fe] (where X = Sc, Ti, and V) as a function of [Fe/H] for different values of the IMF slope parameter, $\alpha_3$. The massive-star yields correspond to the non-rotating case, i.e., $\mathrm{IRV}=0$ km/s. Observational point colours follow the literature subsamples defined in Section~\ref{sec:method}. Blue denotes studies containing both thin- and thick-disc stars, red denotes halo-only samples, black denotes thin-disc-only samples, and green denotes mixed thick-disc/halo samples.}
    \label{fig:alpha3_obs}
\end{figure*}

\begin{figure*}[ht!]
    \centering
    \includegraphics[width=0.8\textwidth]{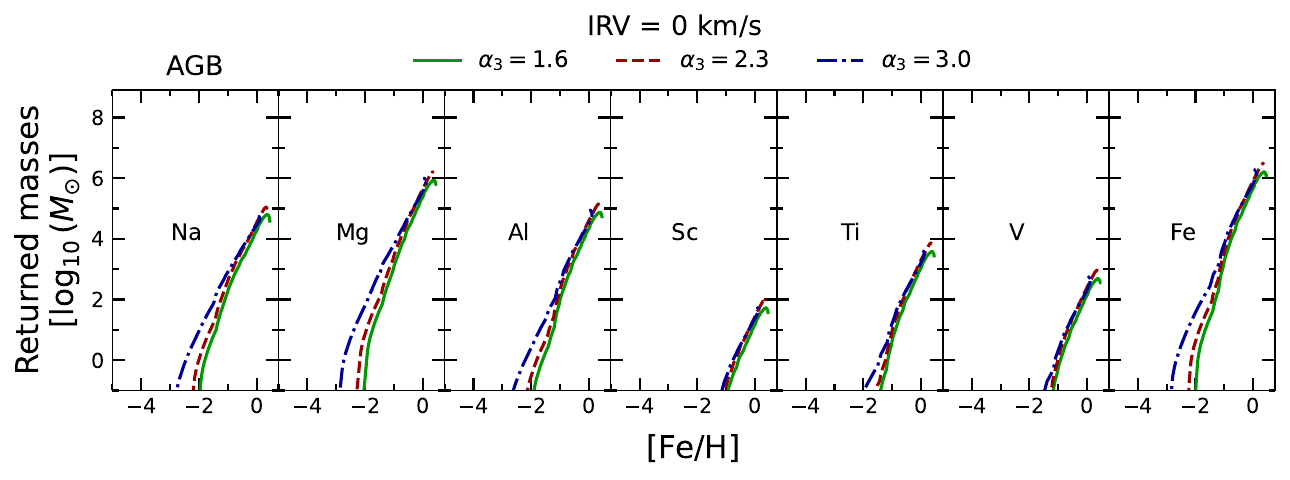}
    \includegraphics[width=0.8\textwidth]{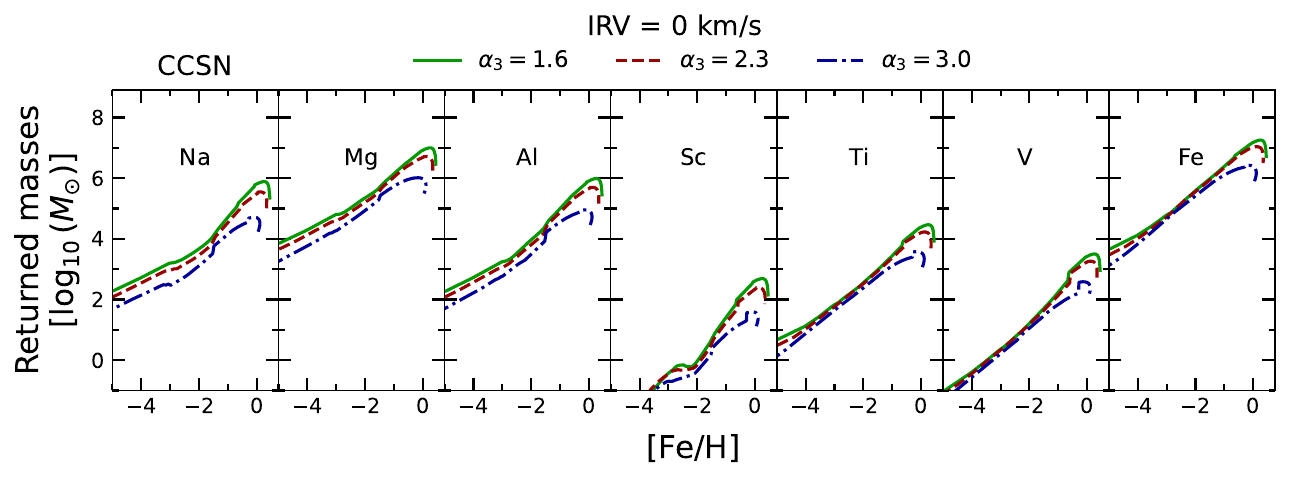}
    \includegraphics[width=0.8\textwidth]{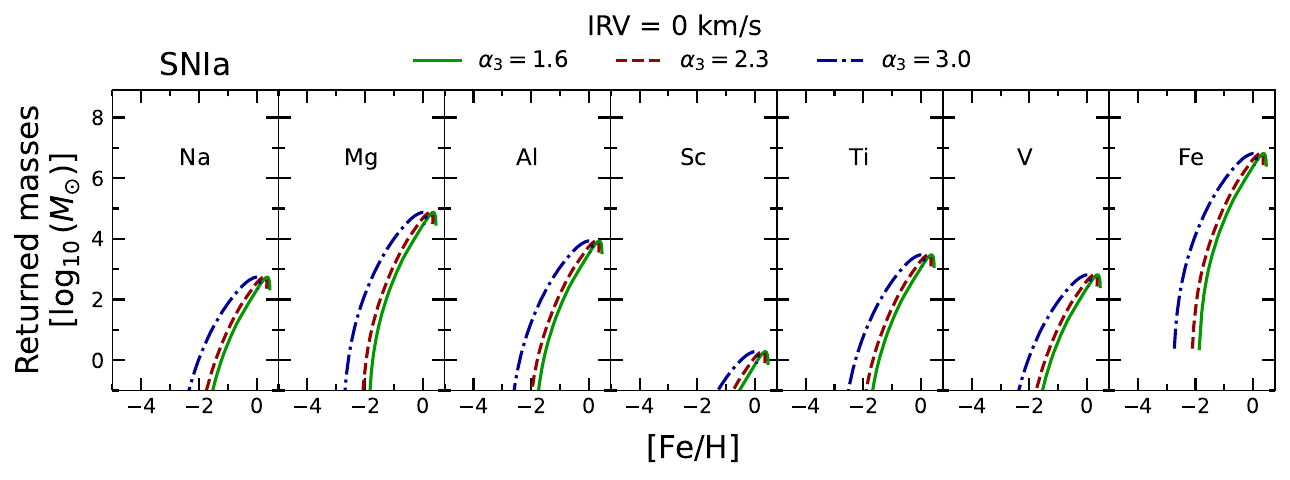}    
    \caption{Returned masses as functions of [Fe/H] for Na, Mg, Al, Sc, Ti, V, and Fe for the three IMF slopes. Here ``returned mass'' denotes the mass of a given element released to the ISM by the indicated source at the model step corresponding to that [Fe/H]. The three stacked panels show the contributions from AGB stars (top), CCSNe (middle), and SNe Ia (bottom). The massive-star yields are the non-rotating set, i.e., $\mathrm{IRV}=0$ km/s, while the AGB and SNe Ia yields are taken from \cite{Cristallo:2015ica} and \cite{Iwamoto:2000as}, respectively.}
    \label{fig:alpha3_iso_return}
\end{figure*}

\begin{figure*}[!ht]
    \centering
    \includegraphics[width=0.8\textwidth]{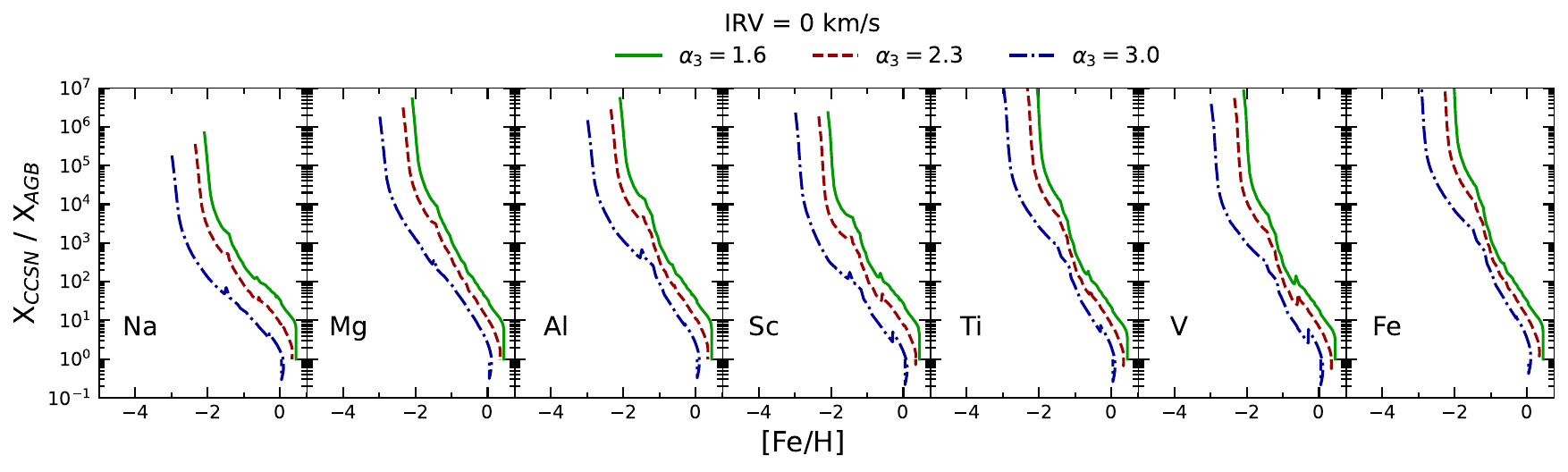}
    \caption{Ratio of the returned mass from CCSNe to that from AGB stars for different high-mass IMF slopes, $\alpha_3=1.6$, 2.3, and 3.0. The massive-star yields are the non-rotating set, i.e., $\mathrm{IRV}=0$ km/s.}
    \label{fig:alpha3_ratio_CCSN_to_AGB}
\end{figure*}

\section{Results}
\label{sec:results}

In this section, we present the main outcomes of our one-zone GCE models and examine how IMF and IRV affect both the global enrichment history and the resulting abundance patterns. We first isolate the effect of the high-mass IMF slope, $\alpha_3$, focusing on how it changes the chemical-enrichment history and the resulting [X/Fe] vs [Fe/H] (X=Sc, Ti, and V) trends through the CCSN and AGB contributions. We then investigate the role of stellar rotation by varying the IRV of massive stars and quantifying how rotation-enhanced CCSN yields modify [Sc/Fe], [Ti/Fe], and [V/Fe], especially at low metallicity. Lastly, we combine IMF-slope variations with rapid rotation (IRV = 300 km/s) to evaluate whether their joint effects can reproduce both the observed [X/Fe] vs [Fe/H] trends and the cross-element correlations among Sc, Ti, and V.

\subsection{Effects of the initial mass function}
\label{subsec:results_imf}
In this subsection, the IMF-slope comparison is carried out with the non-rotating massive-star yield set, i.e., $\mathrm{IRV}=0$ km/s.
Figure~\ref{fig:alpha3_phys_com} shows the effect of varying the high-mass slope of the IMF, $\alpha_3$, on the global evolution of our one-zone GCE. The four panels display, as a function of [Fe/H], the total stellar mass $M_\mathrm{star}$ and the metal-enriched gas mass $M_{Z,g}$, the CCSN and AGB mass-return rates, and the star-formation rate (SFR).

It can be seen from the first panel that the cumulative metal mass $M_{Z,g}$ is clearly affected by the value of $\alpha_3$ around [Fe/H]$=0$. A top-heavy IMF (i.e., lower $\alpha_3$) leads to a greater number of high-mass stars, which produce more metals through nucleosynthesis. As a result, the metal mass is highest for $\alpha_3 = 1.6$ and lowest for $\alpha_3 = 3.0$. At a given metallicity, the top-light IMF gives the highest SFR, while the top-heavy case, $\alpha_3=1.6$, gives the lowest SFR. It is related to the [Fe/H] evolution over cosmic time. The [Fe/H] evolution is fast for the case of a top-heavy IMF due to the metallicity production from many massive stars. Compared to the top-light IMF, the top-heavy IMF shifts to the right side. This trend highlights the critical role of the high-mass IMF slope in determining the overall level of chemical enrichment in galaxies. The total stellar mass $M_\mathrm{star}$ grows in a similar way in all models because the integrated SFR is controlled by the gas mass, and the gas mass evolution is set primarily by the prescribed infall when outflows are neglected.

The second and third panels show the CCSN and AGB mass-return rates as a function of [Fe/H]. At low metallicity ([Fe/H] $\le$ -2), the CCSN mass-return rate is already sensitive to the IMF slope. A top-heavy IMF produces a larger CCSN contribution, while a top-light IMF suppresses it at all [Fe/H]. In contrast, the AGB mass-return rate increases with $\alpha_3$. This opposite trend arises because, in our model, AGB stars span $0.8$--$8\,M_\odot$. When the high-mass slope of the IMF becomes steeper, the number of lower-mass stars grows, so the total number of stars that eventually pass through the AGB phase also increases, leading to a larger AGB contribution for larger $\alpha_3$.

The fourth panel plots the SFR versus [Fe/H]. Differences among the curves mainly reflect how fast each IMF model enriches Fe and shifts the same star-formation history along the [Fe/H] axis.
At very low metallicity, the SFR is highest for the top-light IMF ($\alpha_3 = 3.0$) and lowest for the most top-heavy case ($\alpha_3 = 1.6$), whereas near solar metallicity the ordering is reversed and the $\alpha_3 = 1.6$ model attains the largest SFR. 
This is the same horizontal-shift effect seen in the $M_{Z,g}$ panel.

Figure~\ref{fig:alpha3_obs} shows [X/Fe] for Sc, Ti, and V as a function of [Fe/H], overlaid with observational data from thin-disc, thick-disc, and halo stars. Each curve corresponds to different $\alpha_3$ values. Across all three elements, only small variations with $\alpha_3$ are visible in the predicted [X/Fe] ratios. A more top-heavy IMF (lower $\alpha_3$) tends to produce slightly higher [X/Fe] values for Sc, Ti, and V at low metallicities because of the increased contribution of massive stars. This indicates that while the IMF slope influences the number of massive stars, it alone cannot fully account for the observed abundance patterns of Sc, Ti, and V.

The comparison with observational data shows that none of the IMF slopes explored here fully reproduces the observed trends, except near [Fe/H]$=0$. In the metal-poor regime, the results with all three values of $\alpha_3$ adopted in this study fall below the measured [X/Fe] ratios, which is consistent with previous studies \citep{Gjergo2023,Kobayashi:2020jes,Prantzos2018chemical}. This agreement with \cite{Prantzos2018chemical} is important because the low-metallicity deficit already appears when the Limongi--Chieffi yields are used without strong rotation. The additional contribution of the present paper is to quantify how much that conclusion changes when the IMF high-mass slope and the joint Sc--Ti--V behaviour are examined simultaneously.

Figures~\ref{fig:alpha3_iso_return} and \ref{fig:alpha3_ratio_CCSN_to_AGB} provide additional insight into how the IMF slope changes the contributions of different nucleosynthetic sites, including CCSNe, AGB stars, and SNe Ia. In this paper, returned mass means the mass of a given element that is released into the ISM by a specific enrichment channel at the model step corresponding to the plotted [Fe/H]. Figure~\ref{fig:alpha3_iso_return} displays these returned masses from AGB stars, CCSNe, and SNe Ia as functions of [Fe/H] for the three values of $\alpha_3$. For Sc, Ti, and V, the CCSN contribution dominates in the metal-poor regime ($\mathrm{[Fe/H]} \lesssim -2$), while AGB stars begin to contribute only after the first low- and intermediate-mass stars reach the AGB phase. Iron isotopes receive non-negligible contributions from both CCSN and SNIa, as expected.

Figure~\ref{fig:alpha3_ratio_CCSN_to_AGB} shows $X_\mathrm{CCSN}/X_\mathrm{AGB}$, the ratio of returned mass from CCSN to that from AGB stars, as a function of [Fe/H] for Na, Mg, Al, Sc, Ti, V, and Fe elements.
Because AGB stars start contributing only after a delay, the ratio is not defined at the earliest epochs. As the Galaxy evolves and AGB stars begin to return material to the interstellar medium, the ratio becomes well defined. The top-light IMF slopes suppress the number of massive stars, and therefore the CCSN yields, leading to a decrease in $X_\mathrm{CCSN}/X_\mathrm{AGB}$. This behaviour is further enhanced by the long evolutionary timescales of low-mass AGB stars, whose contributions become significant only at later times. 

Overall, varying the IMF slope shifts the relative CCSN and AGB contributions as expected, while the predicted changes in [Sc/Fe], [Ti/Fe], and [V/Fe] remain limited at low metallicity.

\subsection{Effects of initial rotational velocity}
\label{subsec:results_irv}

\begin{figure*}[ht!]
    \centering
    \includegraphics[width=0.95\textwidth]{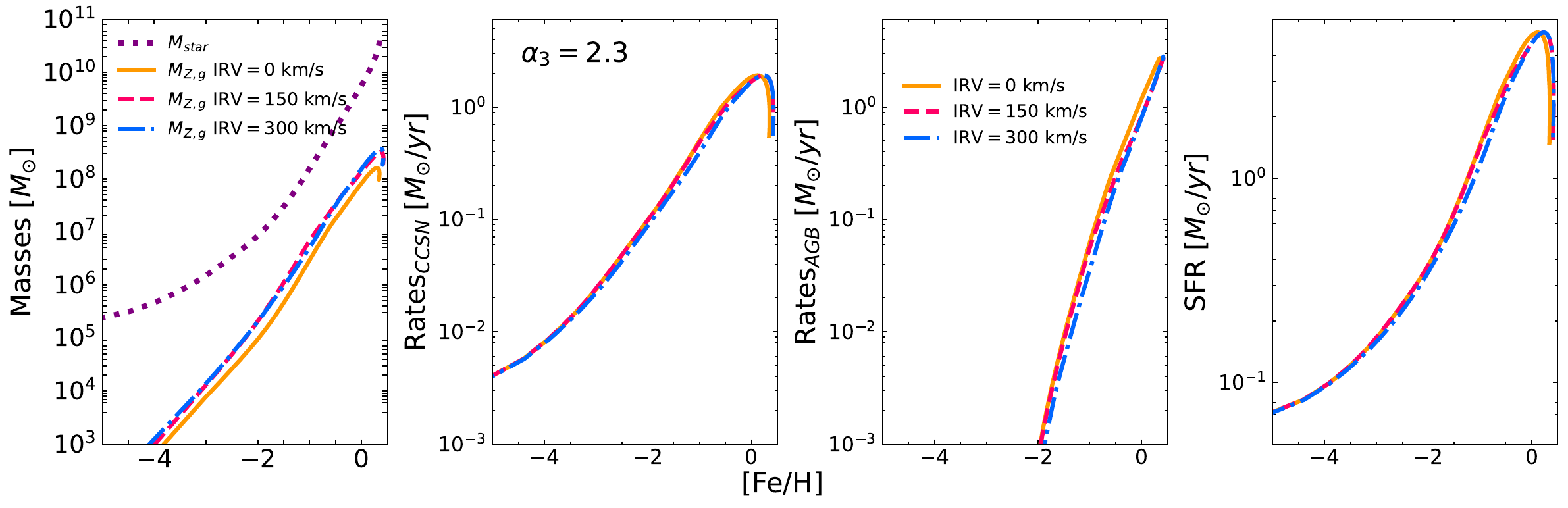}
    \caption{Four-panel comparison of the one-zone GCE histories for three representative initial rotational velocities (IRVs) of the massive-star models. From left to right, the panels show the total stellar mass together with the metal-enriched gas mass, the CCSN mass-return rate, the AGB mass-return rate, and the star-formation rate. In the first panel, the dotted purple line denotes the total stellar mass, $M_\mathrm{star}$. The orange solid, magenta dashed, and blue dot-dashed lines correspond to $\mathrm{IRV}=0$, 150, and 300 km/s, respectively. The mass-return rates and the SFR are expressed in $M_\odot$/yr.}
    \label{fig:irv_phys_com}
\end{figure*}

\begin{figure*}[ht!]
    \centering
    \includegraphics[width=0.95\textwidth]{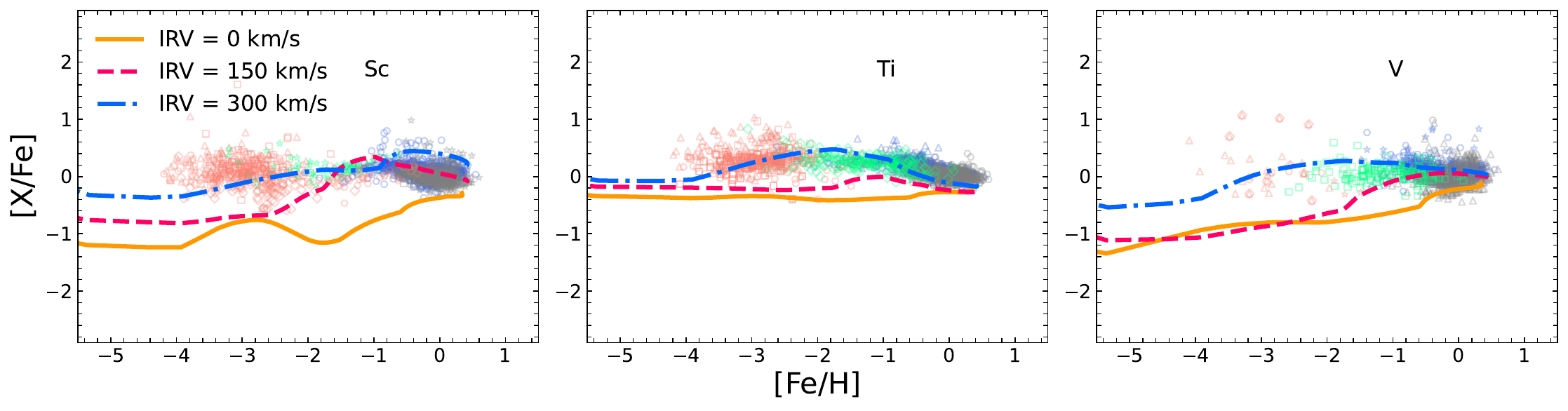}
    \caption{The abundance ratios [X/Fe] (where X = Sc, Ti, and V) as a function of [Fe/H] for different values of IRV. The IMF slope is set at the canonical value, $\alpha_3=2.3$. Observational point colours follow the literature subsamples defined in Section~\ref{sec:method}. Blue denotes studies containing both thin- and thick-disc stars, red denotes halo-only samples, black denotes thin-disc-only samples, and green denotes mixed thick-disc/halo samples.}
    \label{fig:irv_obs}
\end{figure*}

\begin{figure*}[ht!]
    \centering
    \includegraphics[width=0.8\textwidth]{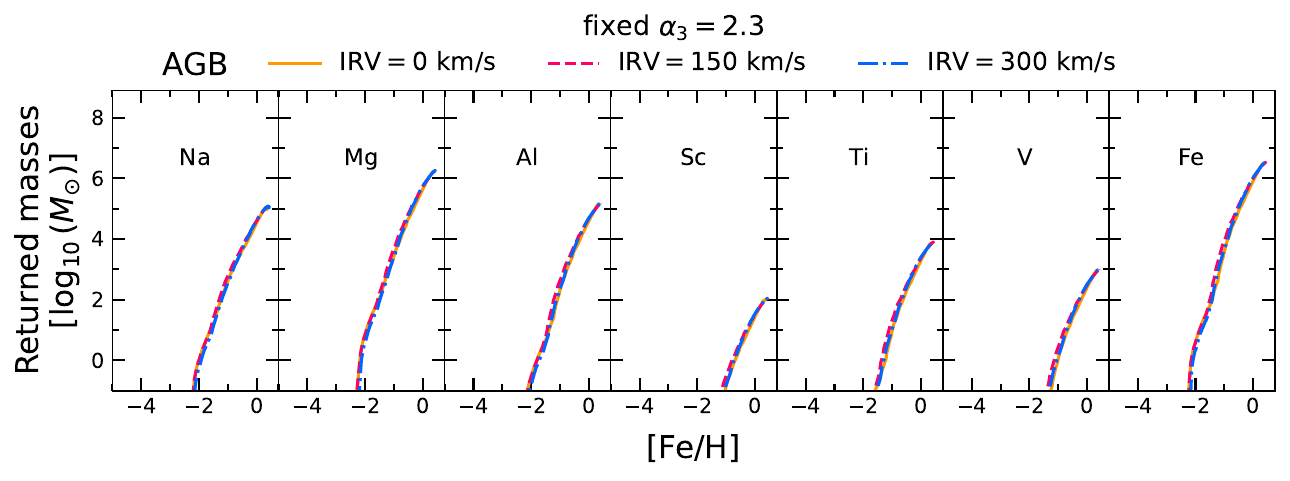}
    \includegraphics[width=0.8\textwidth]{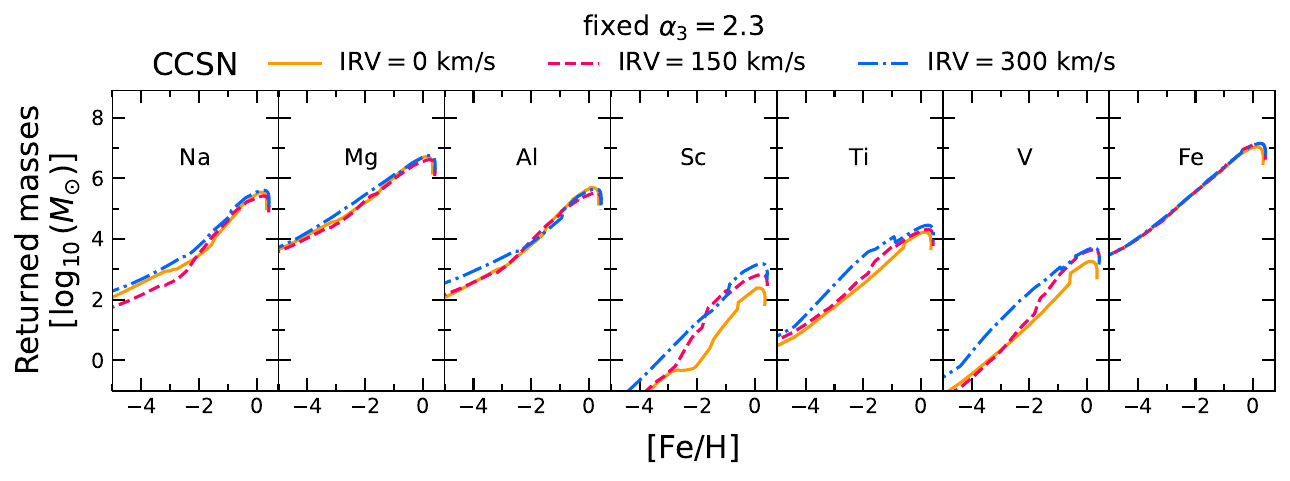}
    \caption{Returned masses as functions of [Fe/H] for Na, Mg, Al, Sc, Ti, V, and Fe for the three representative IRVs, computed at a high-mass IMF slope of $\alpha_3=2.3$. Here ``returned mass'' denotes the mass of a given element released to the ISM by the indicated source at the model step corresponding to that [Fe/H]. The upper and lower panels show the AGB and CCSN contributions, respectively.}
    \label{fig:irv_iso_return}
\end{figure*}

We now investigate the impact of stellar rotation on chemical evolution by varying IRVs of massive stars. In this work, the initial rotational velocity is applied only to the massive stars that contribute to CCSN yields, while AGB yields are taken from non-rotating models. The massive-star yields are taken from \cite{Limongi:2018qgr}, which provide metallicity-dependent nucleosynthesis predictions for stars with IRVs of 0, 150, and 300~km/s.

Figure~\ref{fig:irv_phys_com} shows $M_\mathrm{star}$, $M_{Z,g}$, CCSN and AGB mass-return rates, and the SFR, analogous to Figure~\ref{fig:alpha3_phys_com} but now for different IRVs at set high-mass IMF slope $\alpha_3=2.3$. The total stellar mass $M_\mathrm{star}$ and the SFR follow nearly identical histories in all IRV cases because the underlying star-formation prescription is independent of stellar rotation. In contrast, the metal mass $M_{Z,g}$ increases systematically with IRV because the adopted rotating massive-star models eject larger metal masses. In the \cite{Limongi:2018qgr} grid, this behaviour is linked to rotation-induced mixing between burning regions, which enhances the production of primary $^{14}$N and subsequently $^{22}$Ne and thereby alters the conditions for synthesizing Sc, Ti, and V. Because our GCE calculation adopts the published integrated yield tables, the present comparison shows the combined effect of these rotation-dependent yields rather than separating the individual internal-mixing channels.

The CCSN mass-return rate is only weakly sensitive to IRV because the event frequency is set mainly by the IMF and by the lifetimes of massive stars, which are only modestly affected over the range of rotation velocities considered here. The main IRV dependence instead enters through the CCSN yield amplitudes. Rotating massive-star models return different metal masses to the ISM, which changes the global enrichment history of the one-zone model. Although the AGB yields themselves do not depend on IRV in our model, this IRV-dependent CCSN enrichment slightly changes the subsequent gas composition and evolutionary path, and therefore produces small differences in the AGB returned masses evaluated along the [Fe/H] track.

Figure~\ref{fig:irv_obs} compares the [X/Fe] abundance ratios of Sc, Ti, and V for different IRVs, again overlaid with the observational datasets. As the IRV increases, the predicted [X/Fe] ratios increase across the three elements, particularly at low [Fe/H]. This pattern is a direct consequence of the enhanced yields in rotating stellar models, which synthesize larger amounts of these iron-peak elements.

Among three different values of IRVs, the case with $\mathrm{IRV}=300$ km/s provides the best agreement with the observational data at low metallicity, as shown in Fig.~\ref{fig:irv_obs}, suggesting that rapid rotation in early generations of massive stars can contribute to the observed Sc, Ti, and V data.
The overall decline of the SFR as the Galaxy evolves reduces the formation rate of new massive stars, which further weakens the impact of rotation. 
Thus, changing only the number of massive stars cannot fully reproduce the observed early enhancement of Sc, Ti, and V. 
Our results suggest that rotation-enhanced CCSN yields play a significant role, while the increasing AGB contribution at later times gradually weakens these rotational signatures at higher metallicity ($\mathrm{[Fe/H]} \gtrsim -2$).

\begin{figure*}[!ht]
    \centering    \includegraphics[width=0.8\textwidth]{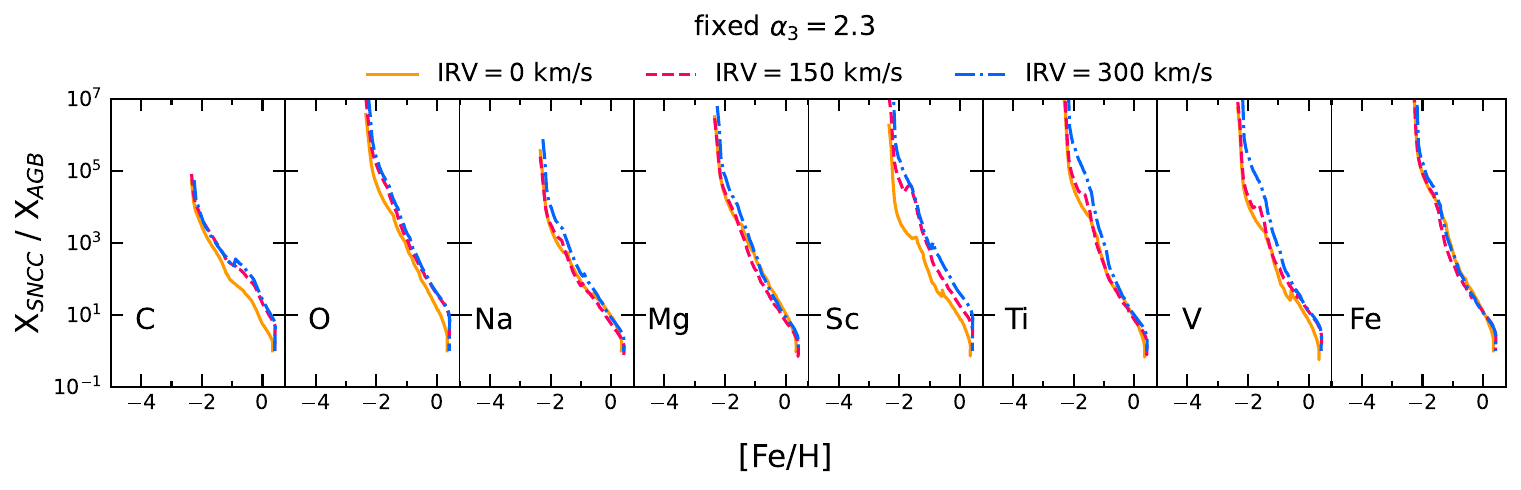}
    \caption{Ratio of the returned mass from CCSN to that from AGB for various IRVs, computed at a value of  $\alpha_3=2.3$.}
    \label{fig:irv_ratio_CCSN_to_AGB}
\end{figure*}

The influence of rotation on the relative contributions from different nucleosynthetic sites is further illustrated in Figures~\ref{fig:irv_iso_return} and \ref{fig:irv_ratio_CCSN_to_AGB}.
Figure~\ref{fig:irv_iso_return} presents the returned masses of elements from AGB stars and CCSN as functions of [Fe/H] for the three IRVs. The comparison is computed at a set value of $\alpha_3=2.3$. For Sc, Ti, and V, the CCSN contribution dominates in the metal-poor regime, while the AGB contribution again becomes important only after $\mathrm{[Fe/H]} \approx -2$. At higher metallicities, AGB stars provide a non-negligible fraction of the total returned mass, reducing the fractional impact of rotation-enhanced CCSN yields. We do not include a separate SNe Ia panel in this figure because the adopted SNe Ia yields and delay-time prescription are fixed across the IRV-only models. Varying the massive-star IRV changes the CCSN contribution but not the SNe Ia yield set.

Figure~\ref{fig:irv_ratio_CCSN_to_AGB} shows the ratio $X_\mathrm{CCSN}/X_\mathrm{AGB}$ as a function of [Fe/H] for the same elements shown in Fig.~\ref{fig:alpha3_ratio_CCSN_to_AGB}. Increased IRV enhances the CCSN yields through rotational mixing, while AGB yields remain unchanged in our model. 
As a result, $X_\mathrm{CCSN}/X_\mathrm{AGB}$ increases with IRV over most of the metallicity range. The increase is notable at very low metallicity ($\mathrm{[Fe/H]} \lesssim -2$), where AGB contributions are still small, especially for Sc, Ti, and V.
This behaviour emphasizes the role of stellar rotation in early chemical enrichment.

\subsection{Combined effect of IMF slope and stellar rotation}
\label{subsec:results_combined}

\begin{figure*}[!ht]
    \centering
    \includegraphics[width=\textwidth]{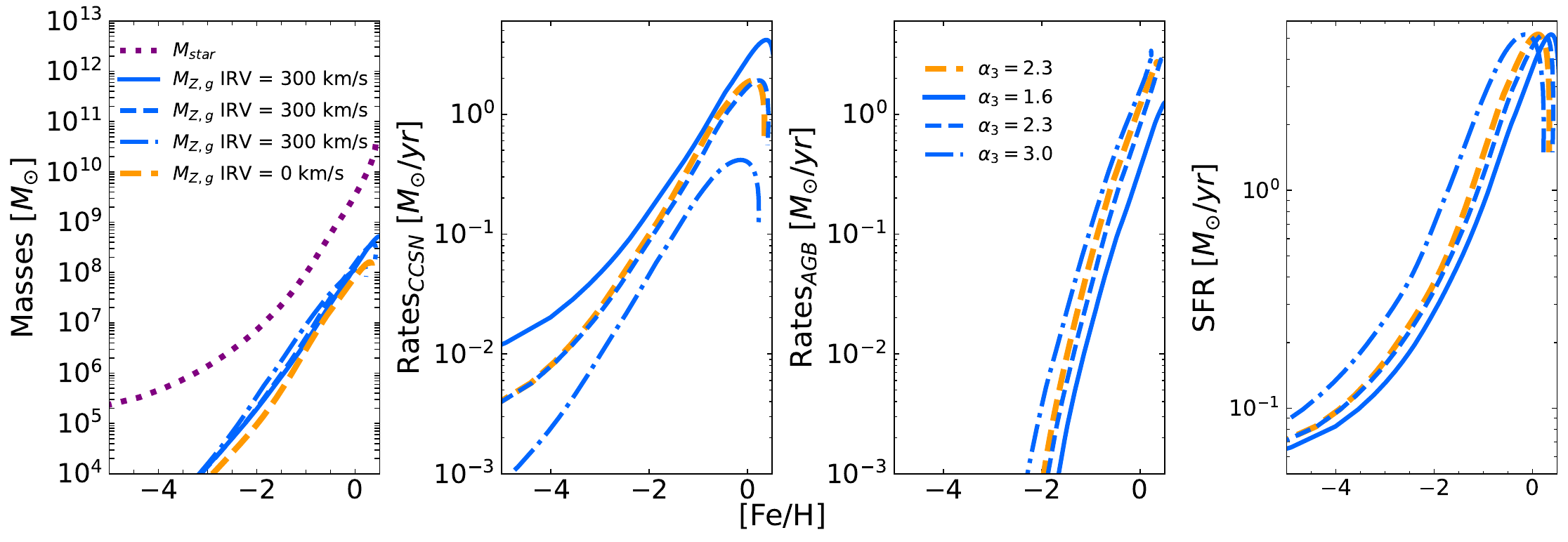}
    \caption{Four-panel comparison of the one-zone GCE histories for three high-mass IMF slopes at an initial rotational velocity $\mathrm{IRV}=300$~km/s. From left to right, the panels show the total stellar mass together with the metal-enriched gas mass, the CCSN mass-return rate, the AGB mass-return rate, and the star-formation rate. In the first panel, the dotted purple line denotes the total stellar mass, $M_\mathrm{star}$. Blue lines show models computed with $\mathrm{IRV}=300$ km/s. Their line styles indicate the IMF slope, with solid for $\alpha_3=1.6$, dashed for $\alpha_3=2.3$, and dot-dashed for $\alpha_3=3.0$. The orange dashed line shows the reference non-rotating model with $\alpha_3=2.3$ and $\mathrm{IRV}=0$ km/s. The mass-return rates and the SFR are expressed in $M_\odot$/yr.}
    \label{fig:irv_300_phys_com}
\end{figure*}

We assess the combined impact of IMF slope and stellar rotation. Figures~\ref{fig:irv_300_phys_com} and \ref{fig:irv_300_obs} present results for models with a fixed IRV of 300 km/s and three different IMF slopes, $\alpha_3 = 1.6$, 2.3, and 3.0. For comparison, a canonical model with $\alpha_3 = 2.3$ and IRV = 0 km/s is also shown.

In Figure~\ref{fig:irv_300_phys_com}, the first panel shows $M_{\rm star}$ and $M_{Z,g}$ as functions of [Fe/H]. 
The growth of $M_{\rm star}$ is almost identical in all cases, while $M_{Z,g}$ depends on the IMF slope.
For models with an IRV of 300 km/s and with the three different IMF slopes, a top-heavy IMF produces a higher $M_{Z,g}$ and therefore the highest curve, while the top-light IMF gives the lowest curve at [Fe/H]$=0$.
However, at the low [Fe/H] region, the top-light IMF has the highest curve. 
It is related to the [Fe/H] evolution over cosmic time. 
The [Fe/H] evolution is fast for the case of a top-heavy IMF due to the metallicity production from many massive stars. 
For $\alpha_3=2.3$, the rotating case yields a slightly higher $M_{Z,g}$ than the canonical model (IRV = 0 km/s). 

The panel for CCSN mass-return rates shows the same IMF ordering. The CCSN contribution is highest for $\alpha_3=1.6$ and lowest for $\alpha_3=3.0$. The AGB panel shows the opposite behavior, because AGB progenitors span 0.8-8~$M_\odot$ and a steeper IMF produces many more low-mass stars that eventually evolve through the AGB phase, so the AGB mass-return rate is larger for higher $\alpha_3$.
The SFR panel displays the corresponding star-formation history as a function of [Fe/H]. 
At a given metallicity, the top-light IMF gives the SFR highest at a given [Fe/H], while the top-heavy case $\alpha_3=1.6$ has the lowest SFR, which means that faster Fe production in the top-heavy IMF shifts a given SFR toward higher [Fe/H].

\begin{figure}[!ht]
    \centering
    \includegraphics[width=0.8\columnwidth]{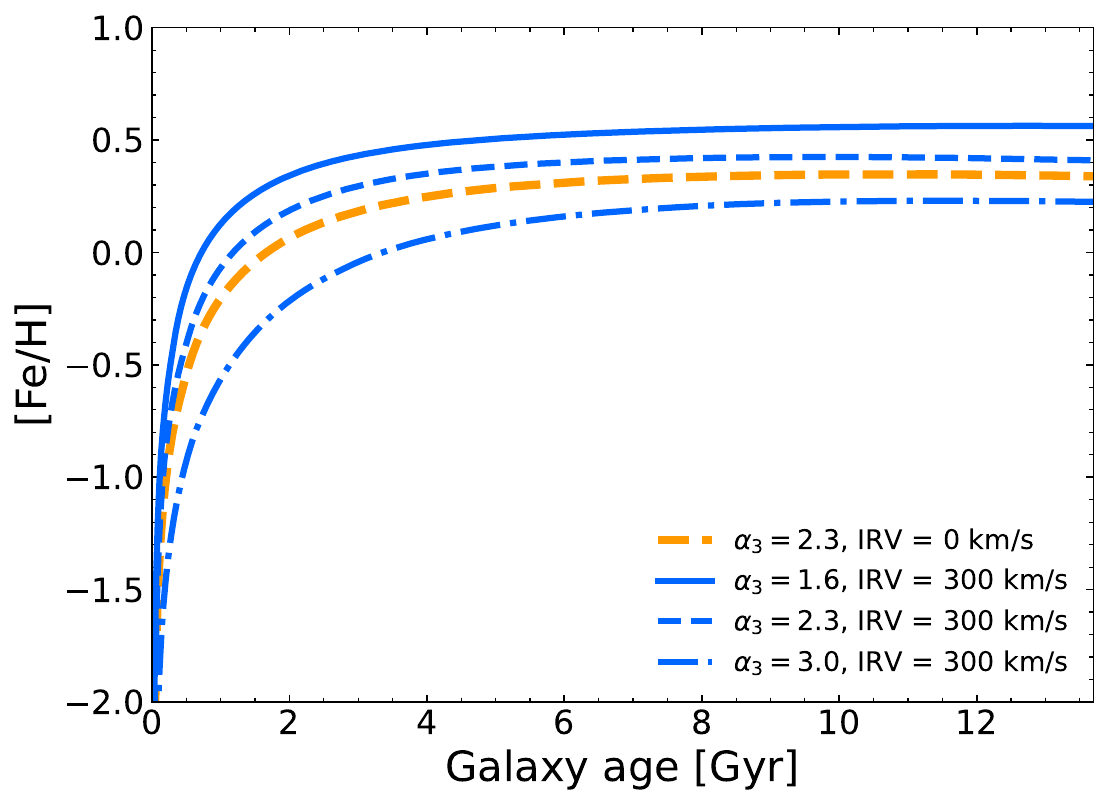}
    \caption{Evolution of [Fe/H] as a function of Galaxy age for the combined-effect models shown in Figure~\ref{fig:irv_300_phys_com}. The orange dashed line shows the reference model with $\alpha_3=2.3$ and $\mathrm{IRV}=0$ km/s. The blue lines show $\mathrm{IRV}=300$ km/s models, with solid, dashed, and dot-dashed lines corresponding to $\alpha_3=1.6$, 2.3, and 3.0, respectively.}
    \label{fig:combined_feh_time}
\end{figure}

\begin{figure*}[!ht]
    \centering
    \includegraphics[width=0.8\textwidth]{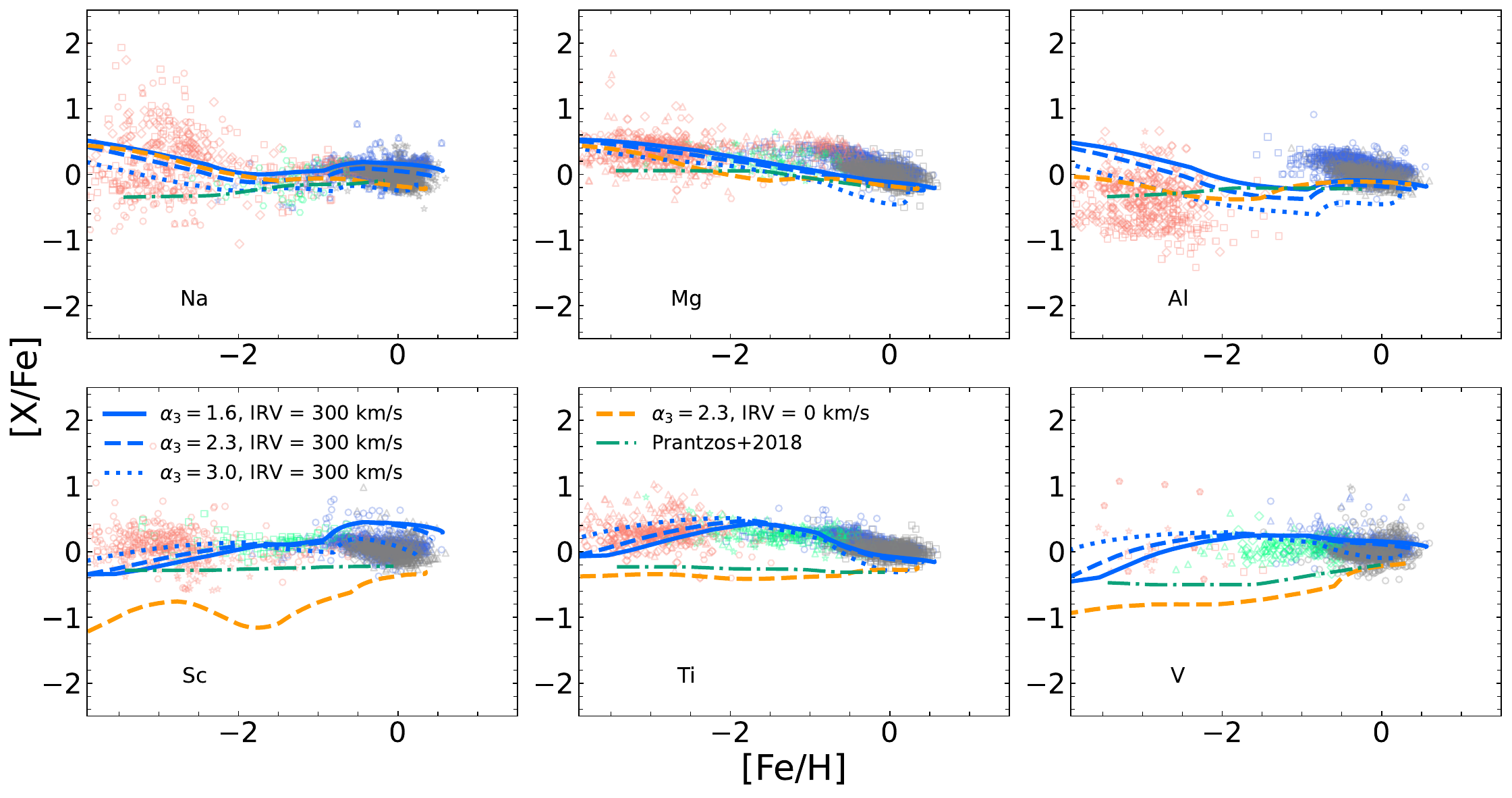}
    \caption{The abundance ratios [X/Fe] (where X = Na, Mg, Al, Sc, Ti, and V) as a function of [Fe/H] for different values of $\alpha_3$ with $\mathrm{IRV}=300$ km/s. Observational point colours follow the literature subsamples defined in Section~\ref{sec:method}. Blue denotes studies containing both thin- and thick-disc stars, red denotes halo-only samples, black denotes thin-disc-only samples, and green denotes mixed thick-disc/halo samples. The green dot-dashed line is taken from \citet{Prantzos2018chemical}, their Fig.~13.}
    \label{fig:irv_300_obs}
\end{figure*}

A comparison of the first panel of Figure~\ref{fig:irv_300_phys_com} between the $\mathrm{IRV}=0$ and $\mathrm{IRV}=300$ km/s cases when the high-mass IMF slope is $\alpha_3=2.3$ shows that stellar rotation mainly increases the build-up of metal-enriched gas because the rotating massive-star models eject larger metal masses. The ordering of the tracks in Figure~\ref{fig:irv_300_phys_com} should therefore be interpreted as a comparison at the same [Fe/H], not as a direct time sequence. The corresponding time-domain view is shown in Figure~\ref{fig:combined_feh_time}. At a given evolutionary time, the rotating model reaches slightly higher [Fe/H] than the non-rotating reference model. The detailed shapes of the CCSN, AGB, and SFR curves as functions of [Fe/H] are still determined mainly by $\alpha_3$, which sets the numbers of massive and low-mass stars that form.
Figure~\ref{fig:combined_feh_time} shows this time dependence directly by plotting [Fe/H] as a function of Galaxy age for the same combined-effect models used in Figure~\ref{fig:irv_300_phys_com}. This diagnostic plot confirms that the rapidly rotating models reach a given [Fe/H] slightly earlier than the non-rotating reference model, while changes in $\alpha_3$ for the same $\mathrm{IRV}=300$ km/s introduce only modest differences in the age--metallicity track over the displayed range. We use this figure only to clarify the mapping between evolutionary time and [Fe/H]. The abundance-ratio comparisons below remain the primary constraints of this study.

Figure~\ref{fig:irv_300_obs} presents the abundance ratios [X/Fe] for Na, Mg, Al, Sc, Ti, and V as a function of [Fe/H] for different IMF slopes with a fixed IRV of 300 km/s. For Na, Mg, and Al, the three rotating models give very similar [X/Fe] at the lowest metallicities ($\mathrm{[Fe/H]} \lesssim -4$). As [Fe/H] increases, the curves gradually spread out and the differences among the IMF slopes become more visible.
In this metallicity range, the top-heavy IMF ($\alpha_3=1.6$) produces the highest [X/Fe], while the steep IMF ($\alpha_3=3.0$) gives the lowest values.
Therefore, [Na/Fe], [Mg/Fe], and [Al/Fe] decrease as $\alpha_3$ increases.
Sc, Ti, and V behave differently.
At low metallicity ([Fe/H] $\lesssim -2$), the ordering among the rotating models is reversed. The model with $\alpha_3=3.0$ gives the largest enhancement, $\alpha_3=2.3$ is intermediate, and $\alpha_3=1.6$ gives the lowest [X/Fe].
At higher metallicity ([Fe/H] $\gtrsim -1$), the three rotating curves for Sc, Ti, and V converge and remain within a narrow band.
In contrast, the canonical model stays clearly below the rotating cases across the full metallicity range.

To quantify the visual comparison in Figure~\ref{fig:irv_300_obs}, we binned the physical-neutral observational sample in metallicity using set bins of width 0.25 dex in [Fe/H], excluded bins containing fewer than five stars, and adopted the median [X/Fe] in each retained bin as the representative observational trend. For the $i$th retained bin, we defined the residual as
\begin{equation}
    r_i = [X/\mathrm{Fe}]^{\mathrm{model}}_i - [X/\mathrm{Fe}]^{\mathrm{obs,\,med}}_i ,
\end{equation}
and quantified the goodness of agreement by the root-mean-square residual
\begin{equation}
    \mathrm{RMS} = \sqrt{\frac{1}{N}\sum_{i=1}^{N} r_i^2},
\end{equation}
where $N$ is the number of retained metallicity bins for each element. In Table~\ref{tab:binned_rms_total}, we report the mean RMS, defined as the average of the element-by-element RMS values over the six elements Na, Mg, Al, Sc, Ti, and V. Among the four models considered here, the rotating model with $\alpha_3 = 3.0$ and IRV = 300 km/s gives the smallest mean RMS, closely followed by the rotating model with $\alpha_3 = 2.3$ and IRV = 300 km/s, whereas the non-rotating canonical model with $\alpha_3 = 2.3$ and IRV = 0 km/s shows the weakest agreement overall.

\begin{table*}[ht!]
    \centering
    \caption{Binned RMS comparison between the model tracks and the physical-neutral observational sample.}
    \label{tab:binned_rms_total}
    \begin{tabular}{lrrr}
        \toprule
        Model & Mean RMS & Total bins used & $N_{\mathrm{elem}}$ \\
        \midrule
        $\alpha_3 = 3.0$, IRV = 300 km/s & 0.2430 & 105 & 6 \\
        $\alpha_3 = 2.3$, IRV = 300 km/s & 0.2458 & 105 & 6 \\
        $\alpha_3 = 1.6$, IRV = 300 km/s & 0.2774 & 105 & 6 \\
        $\alpha_3 = 2.3$, IRV = 0 km/s   & 0.5249 & 105 & 6 \\
        \bottomrule
    \end{tabular}
\end{table*}

\begin{figure*}[!ht]
    \centering
    \includegraphics[width=\textwidth]{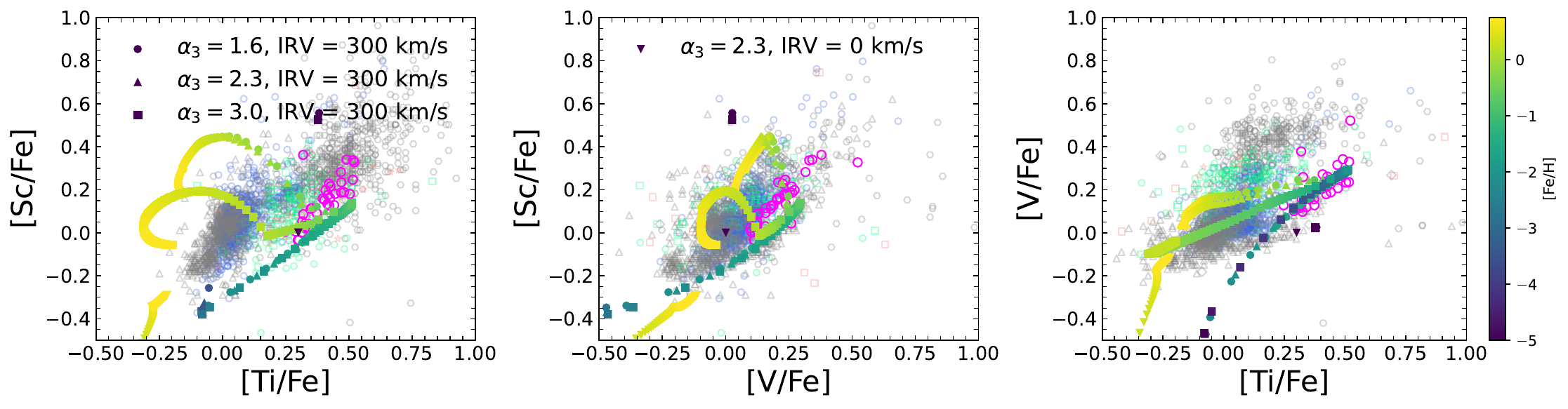}
    \caption{Correlations among the observed abundances of Sc, Ti, and V. The left, middle, and right panels show [Sc/Fe] versus [Ti/Fe], [Sc/Fe] versus [V/Fe], and [V/Fe] versus [Ti/Fe], respectively. Observational data are color-coded by literature subsample, with blue for thick- and thin-disc samples, red for halo samples, black for thin-disc samples, and green for mixed thick-disc and halo samples. The full reference list is shown in Table~\ref{tab:elemobs}. Magenta points are taken from \cite{Sneden2023_IronPeakWPMS}, which is a selection of warm very metal-poor Galactic stars observed with the Keck HIRES spectrograph. The evolutionary model tracks are shown as filled markers, where the shapes correspond to the legend. Tracks outlined in blue represent variations in IMF slope computed using massive-star yields with $\mathrm{IRV}=300$~km/s. Tracks outlined in yellow refer to the canonical IMF slope, $\alpha_3=2.3$, with $\mathrm{IRV}=0$~km/s. The color within each marker indicates the [Fe/H] abundance, as shown in the colorbar on the right.
    }
    \label{fig:irv_300_xfe_yfe}
\end{figure*}

To further examine the abundance correlations among Sc, Ti, and V under given IRV conditions, Figure~\ref{fig:irv_300_xfe_yfe} shows the correlations of [Sc/Fe] versus [Ti/Fe], [Sc/Fe] versus [V/Fe], and [V/Fe] versus [Ti/Fe]. We also include additional Sc–Ti–V measurements from \cite{Sneden2023_IronPeakWPMS} (magenta), which provide more data for comparing the abundance correlations. Model calculations are plotted for the three IMF slopes at IRV = 300~km/s, together with the reference model having $\alpha_3 = 2.3$ and IRV = 0~km/s. Observational data are colour-coded by Galactic component, and model points are colour-coded by [Fe/H].

These correlation plots complement the [X/Fe]–[Fe/H] trends by visualizing the co-evolutions of the iron-peak elements i.e., Sc, Ti, and V. 
For top-heavy IMF slopes ($\alpha_3 = 1.6$) with IRV = 300~km/s, the model results move rapidly from low to high [Fe/H], producing strong low-metallicity enhancement in [Sc/Fe], [Ti/Fe], and [V/Fe] that follows the overall observed sequences. On the other hand, the canonical model appears only at higher metallicity, as expected when the enrichment track reaches a given [Fe/H] after more evolutionary time.
This indicates that stellar rotation makes the model reach the observed iron-peak-enhanced regime at earlier evolutionary time and lower metallicity, thereby contributing to the enrichment pattern traced by these elements.

The [Sc/Fe]–[Ti/Fe] plane shows a clear correlation that agrees well with observations across all the IMF slopes. The [Sc/Fe]–[V/Fe] and [V/Fe]–[Ti/Fe] planes likewise indicate that the co-production of V with Sc and Ti is enhanced in models with both rapid stellar rotation and a moderately top-heavy IMF. Notably, only the IRV = 300 km/s models reproduce the iron-peak-enhanced metal-poor stars at [Fe/H]$\approx-2$ (magenta symbols), which supports the idea that fast rotation in early generations of massive stars is required. 
Overall, the combined effects of IMF slope and stellar rotation bring the results closer to both the observed individual [X/Fe]–[Fe/H] relations of Sc, Ti, and V, and their internal correlations seen in Galactic stars.

\section{Summary and Discussion}
In this study, we investigated the Galactic chemical evolution of Sc, Ti, and V by examining how two key stellar-population parameters influence their nucleosynthetic histories: the high-mass slope of the initial mass function (IMF) and the initial rotational velocity (IRV) of massive stars. Using a one-zone GCE framework \citep{Gjergo2023}, we studied the dependence of these three iron-peak elements on the changes in both parameters. We compared our results with an 
extensive compilation of observational data spanning the Galactic halo, thick disc, and thin disc.

The IMF slope has been inferred to vary with metallicity in the solar neighborhood \citep{Li+2023}. This motivates testing whether such variation affects the abundance trends inferred from GCE, which for Sc, Ti, and V are commonly under-produced.
We found that variations in the IMF slope affect the absolute level of metal enrichment by changing the number of CCSN progenitors.
A top-heavy IMF (a flatter high-mass slope, such as $\alpha_3 = 1.6$) enhances the absolute production of Sc, Ti, and V in CCSNe (Fig.~\ref{fig:alpha3_iso_return}), but the [X/Fe] ratios depend weakly on $\alpha_3$ because iron yields increase proportionally. Consequently, adjusting the IMF slope alone cannot account for the low-metallicity overabundances of these elements observed in metal-poor stars.

Stellar rotation has a much stronger impact. 
High IRV models (300 km/s) enhance the absolute mass of Sc, Ti, and V returned by CCSNe, leaving iron unaffected (Fig.~\ref{fig:irv_iso_return}).
This results in the enhanced [X/Fe] ratios observed at low metallicities ($\mathrm{[Fe/H]} < -2$). 
Rotation-induced mixing alters the pre-supernova structure and composition, and in \cite{Limongi:2018qgr}, the yield grid leads to enhanced Sc, Ti, and V yields at low metallicity, which in turn improves the agreement with observational data of halo-star abundances.

When stellar rotation is included together with varying the high-mass slope of IMF, the results agree better with the observed [X/Fe] for Sc, Ti, and V over the metallicity range spanning from halo to disc stars. Variations of the high-mass IMF slope mainly modulate the relative weighting of CCSN ejecta and the pace at which the model reaches a given [Fe/H]. This produces secondary shifts in [X/Fe] at a given [Fe/H]. 
In the yield set of \cite{Limongi:2018qgr}, rotating massive-star models produce higher Sc, Ti, and V yields relative to Fe compared to the non-rotating models. This effect dominates the change in the inferred abundance tracks.
Overall, stellar rotation is the primary driver of the improved agreement with the data in our results, while IMF variations provide a secondary modulation.

However, some discrepancies remain, most notably at higher [Fe/H], where Sc may be overestimated. 
One-zone models treat the interstellar medium as instantaneously mixed and spatially uniform by construction, so they cannot address the impact of dynamics on the connection between halo, thick-disc, and thin-disc populations. 
In addition, our models assume different but invariant IMF slopes and treat each IRV as a representative velocity rather than as a metallicity-dependent distribution of birth spins. This simplified setup is useful for isolating the sensitivity to the published rotating-yield sets, although a realistic distribution of initial stellar rotation rates remains an important extension for future work.
The time-domain diagnostic in Figure~\ref{fig:combined_feh_time} also clarifies when a statement refers to evolutionary time and when it refers to position along the [Fe/H] track. Because the present paper focuses on abundance ratios of Sc, Ti, and V rather than a full calibration of the Galactic age--metallicity relation, we do not use this diagnostic as an additional fitting constraint.
Future work should implement an environment-dependent IMF, as supported by independent observational constraints and discussed by \cite{Kroupa+2026} and \cite{gjergo2025mass}.
Lastly, limitations may originate from the current CCSN yield prescription, including the treatment of neutrino-processed ejecta \citep{Sieverding+2023}, explosion energy and fallback, including hypernova-like events \citep{NOMOTO2006424}, mass cuts and remnant formation in the progenitor core structure that determine when or whether a supernova occurs \citep{Sukhbold+2016, Limongi:2018qgr}, multi-dimensional mixing \citep{Sieverding+2023}, and nuclear-rate sensitivities in the $\alpha$-rich freeze-out and explosive Si-burning regime \citep{Magkotsios+2010,Hermansen+2020}. In the iron-rich regime, the tension may also be affected by uncertainty in the SNe Ia channels \citep{Gronow+2021}.

Overall, it is central at present to incorporate rotation in massive-star yields to reproduce the observed Sc, Ti and V trends. The existing IMF variations in the Milky Way disc introduce a minor but non-negligible modulation. These results motivate joint tests of massive-star physics and early star-formation conditions to test abundance patterns across Galactic populations.

\section*{Acknowledgements}
We thank the anonymous referee for the helpful feedback which greatly improved the quality of this manuscript. 
This work was supported in part by the Institute for Basic Science, Korea (IBS-R031-D1) and by the National Research Foundation of Korea funded by Ministry of Science and ICT (RS-2024-00436392).
E.G. acknowledges the support of the National Natural Science Foundation of China (NSFC) under grants Nos. 1251101411, 12533003, 1257030642.
E.G. acknowledges the Program for Innovative Talents, Entrepreneur in Jiangsu.
T.K. acknowledges the support in part by the National Key R\&D Program of China (2022YFA1602401) and the National Natural Science Foundation of China (No. 12335009 \& 12435010).

\bibliography{ref}{}
\bibliographystyle{aasjournalv7}

\end{document}